\documentclass{cernyrep}
\usepackage{texnames}
\usepackage[T1]{fontenc}
\usepackage{subfigure}
\usepackage{color}					
\usepackage{amssymb}
\newcounter{query}

\renewcommand{\rmd}{{\rm d}}

\title{Beam Dynamics in Synchrotrons}

\author{B.J.~Holzer}
\institute{CERN, Geneva, Switzerland}

\begin{document}
\maketitle

\begin{abstract}
This paper gives an overview of particle dynamics in synchrotrons and storage rings. Both the transverse and the longitudinal plane are described in a linear approximation. The main emphasis is on giving an introduction to the basic concepts and allowing the reader to deduce the main parameters of a machine, based on some simple scaling laws.

\end{abstract}

\section{Introduction}

We would like to start this little overview with some kind of definition of a synchrotron, in an attempt to achieve the impossible task of
summarizing in a few lines the key issues associated with such a machine. And we would like to ask you, the esteemed reader, to come back to this
point at the end of the story and let us know whether or not the definition is a valuable one.

So, a synchrotron is a type of circular accelerator that needs:
\begin{itemize}
\item a magnetic bending field to keep the particles on a closed, more or less circular orbit;
\item a mechanism to lock this $B$-field to the changing particle energy and thus keep the particles
on, or close to, this design orbit over the complete energy range of the machine;
\item focusing forces  that follow the energy gain of the beam to keep the particles together, and that ultimately lead to a
 well-defined beam size;
\item a Radio Frequency (RF) structure to accelerate the particles and create the necessary energy gain per turn via longitudinal electric fields;
\item a mechanism to synchronize the RF frequency to the timing of the circulating particles and to provide a longitudinal focusing (phase-focusing) effect
that keeps the particles longitudinally bunched.
\end{itemize}

So much for the definition.

This seems to deserve a remark to reassure the reader: these machines exist, they are very robust, they deliver stable particle beams, and, most importantly, they can be built.

\begin{figure}[ht]
\begin{center}
\includegraphics[width=0.5\columnwidth]{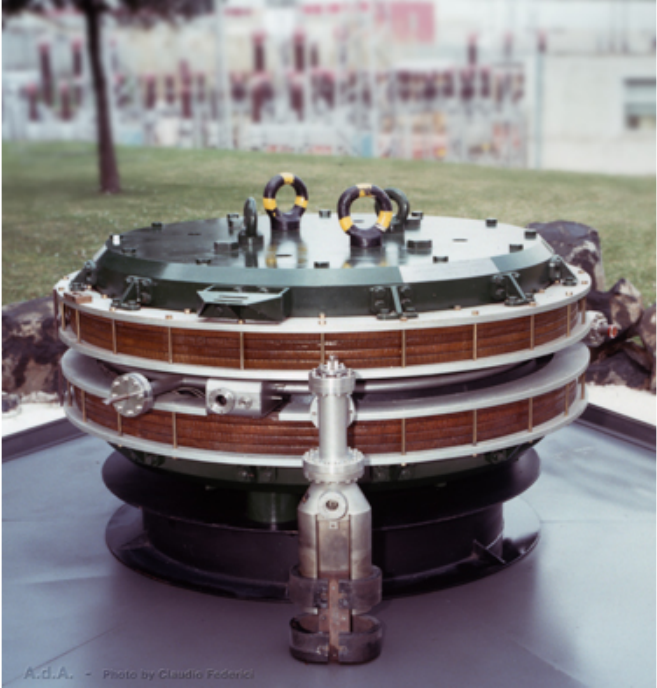}
\caption{ADA, the first electron positron collider ring}
\label{ADA}
\end{center}
\end{figure}

\begin{figure}[ht]
\begin{center}
\includegraphics[width=0.80\columnwidth]{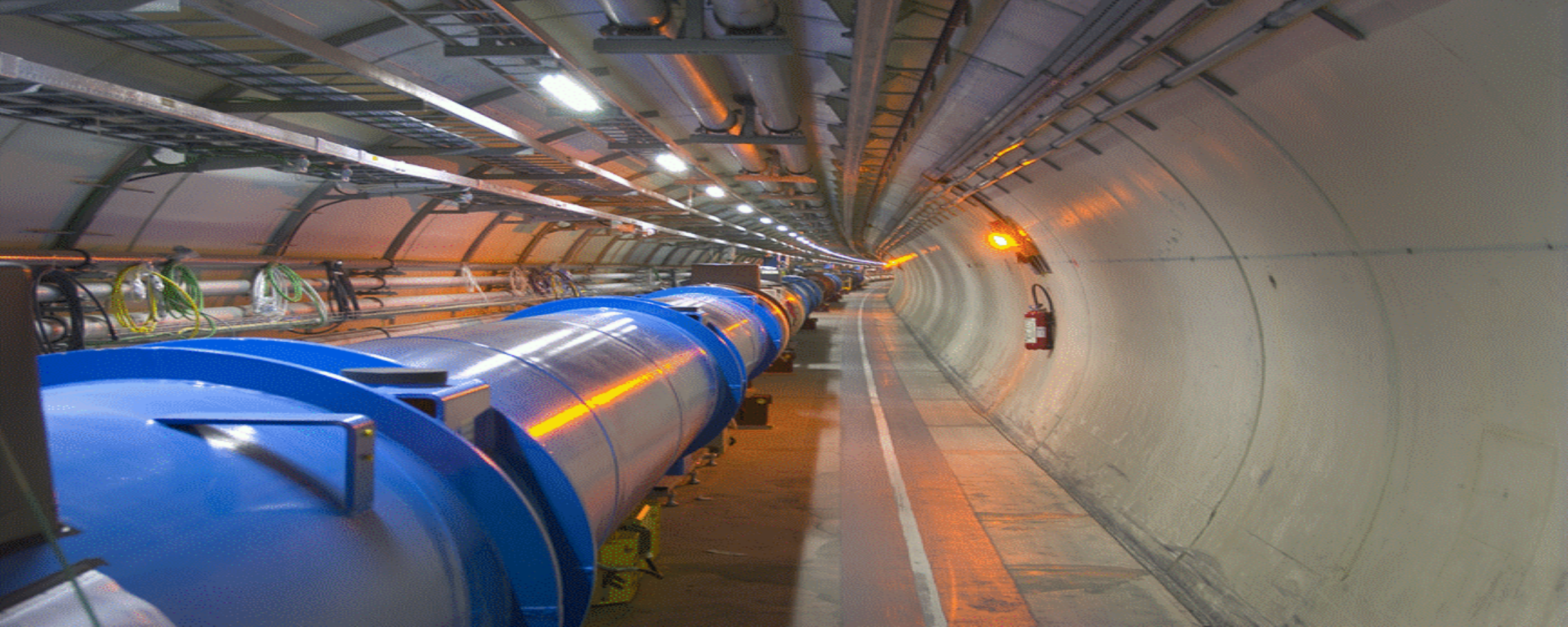}
\caption{A view of the tunnel of the LHC proton--proton collider at CERN, Geneva}
\label{lhc_tunnel}
\end{center}
\end{figure}

\begin{figure}[ht]
\begin{center}
\includegraphics[width=0.60\columnwidth]{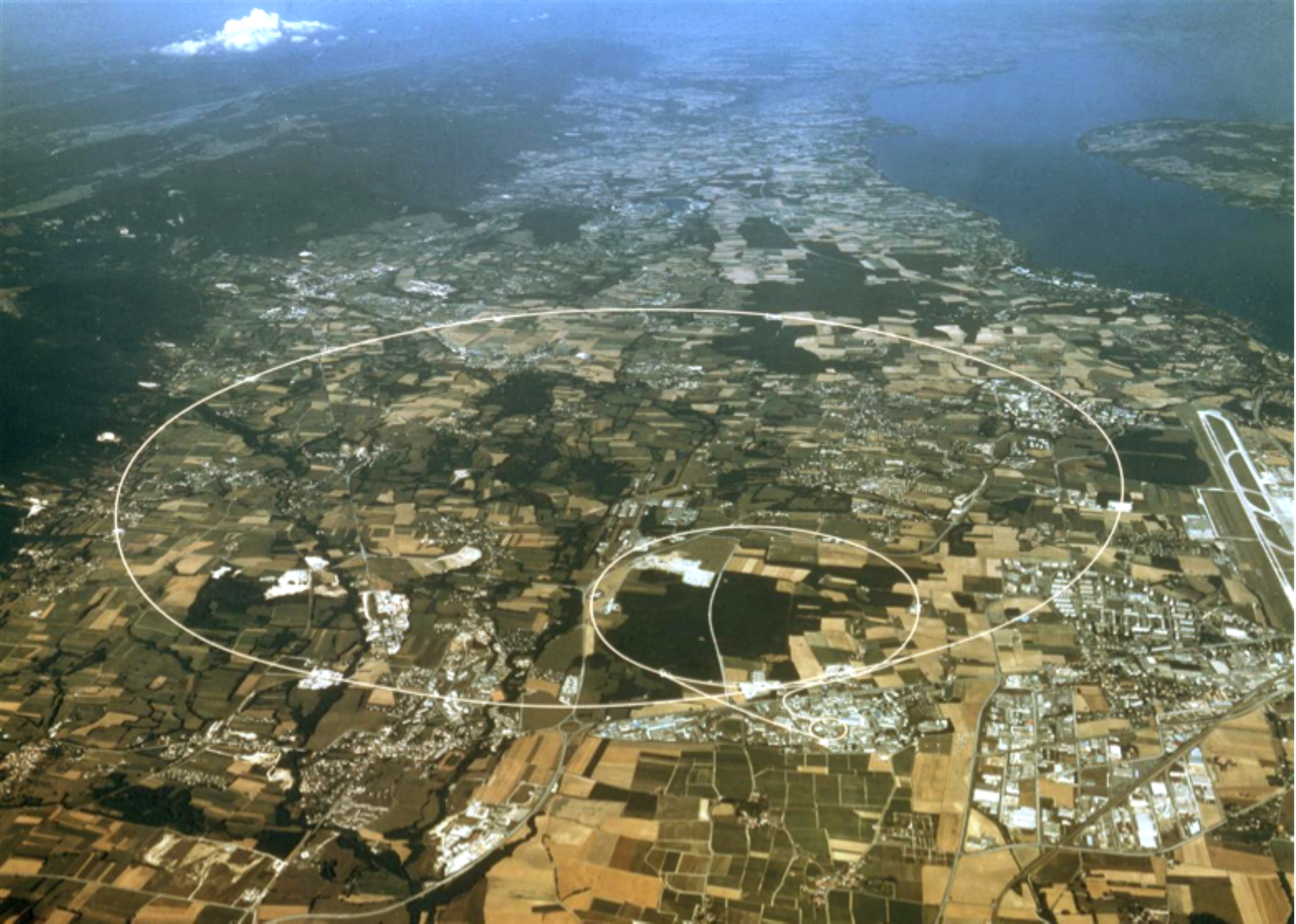}
\caption{The LHC proton--proton collider}
\label{lhc}
\end{center}
\end{figure}

Two examples will act as proof of this strong statement:
the ADA (Annelli de Accumulatione) (Fig.~\ref{ADA}), as far as we know, the very first particle collider and certainly one of the smallest synchrotrons, built in Frascati by Bruno Touschek in 1944;
and the LHC \cite{LHC}, at present the largest storage ring ever built, running at the highest achievable particle energies at CERN (Figs.~\ref{lhc_tunnel} and \ref{lhc}).

\section{Introduction to transverse beam dynamics}
The transverse beam dynamics of charged particles in an accelerator describes the movement of single particles under the influence of the external transverse bending and focusing fields. It includes the detailed arrangement (for example, their positions in the machine and their strength) of the accelerator magnets used to obtain well-defined, predictable parameters of the stored particle beam, and it describes methods to optimize the trajectories of single particles, as well as the dimensions of the beam considered as an ensemble of many particles. A detailed treatment of this field in full mathematical detail, including sophisticated lattice optimizations such as the right choice of the basic lattice cells and the design of dispersion suppressors or chromaticity compensation schemes, is beyond of the scope of this basic overview. For further reading and for more detailed descriptions, we therefore refer to the more detailed explanations in \cite{LHC,Wille,floquet,bjh_erice}.

\subsection{Geometry of the ring}

In general, magnetic fields are used in circular accelerators to provide the bending force and to focus the particle beam. In principle, the use of electrostatic fields would be possible as well, but at high momenta (i.e., if the particle velocity is close to the speed of light), magnetic fields are much more efficient. The force acting on the particles, the Lorentz force, is given by
\begin{equation}
\mathbf{F} = q \cdot (\mathbf{E}+ \mathbf{v} \times \mathbf{B}).
\label{lorentz}
\end{equation}
For high-energy particle beams, the velocity $v$ is close to the speed of light and so represents a nice amplification factor whenever we apply a magnetic field. As a consequence, it is much more convenient to use magnetic fields for bending and focusing the particles.

Therefore, neglecting electric fields for the moment, we write the Lorentz force and the centrifugal force on the particle on its circular path as
\begin{align}
F_{\rm Lorentz}&=e \cdot v \cdot B,\\
F_{\rm centrifugal}&=\frac{\gamma m_{0} v^{2}}{\rho}.
\end{align}
Assuming an idealized homogeneous dipole magnet along the particle orbit, having pure vertical field lines, we define the condition for a perfect circular orbit as equality between these two forces. This yields the following condition for the idealized ring:
\begin{equation}
\frac{p}{e}={B \cdot \rho},
\end{equation}
where we are referring to protons and have accordingly set $q = e$.
This condition relates the so-called beam rigidity $B\rho$ to the momentum of a particle that can be carried in the storage ring, and it ultimately defines, for a given magnetic field of the dipole magnets, the size of the storage ring.

\begin{figure}[h]
\begin{center}
\includegraphics[width=0.6\columnwidth]{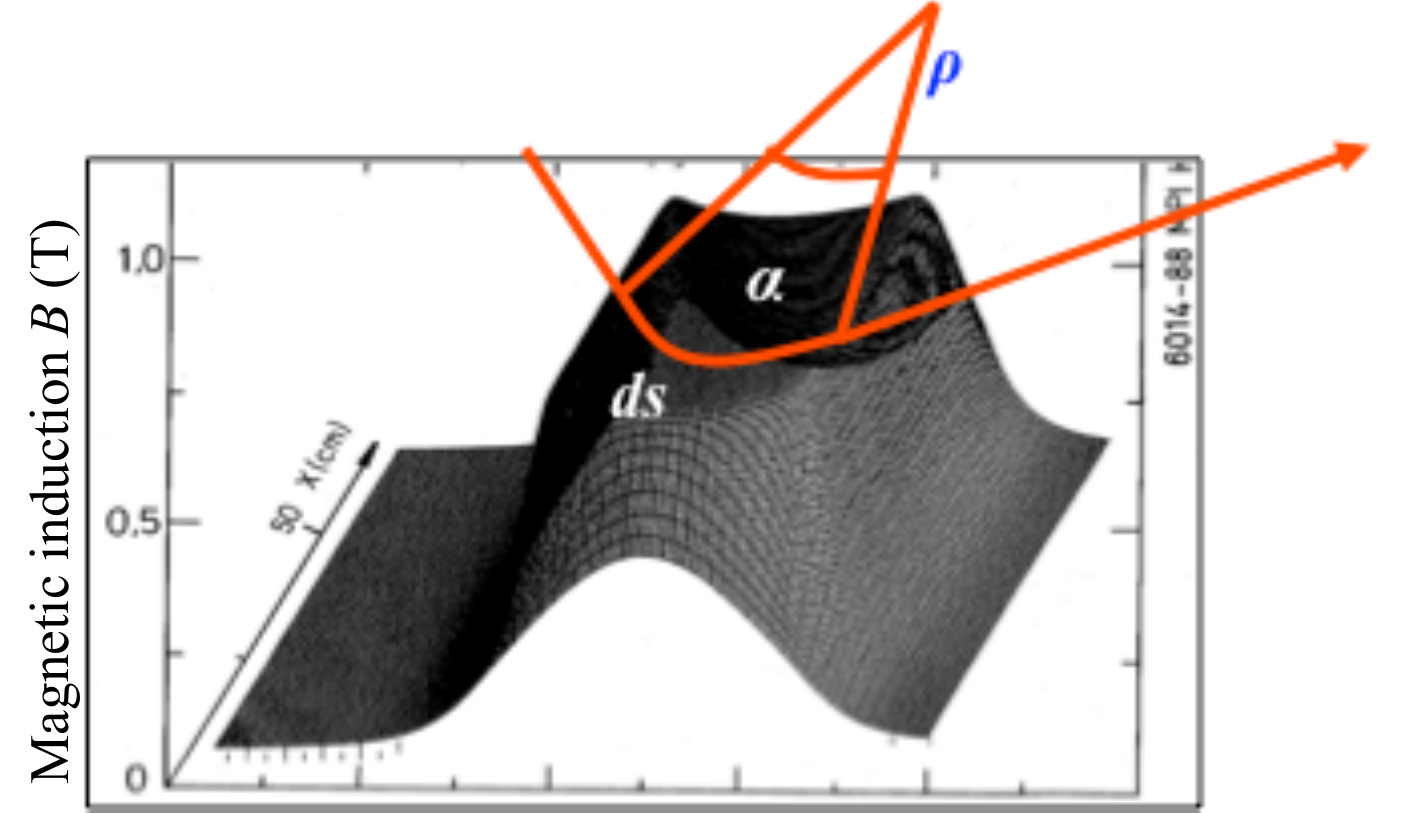}
\caption{Field map of a storage ring dipole magnet, and schematic path of a particle}
\label{TSR_dipole_field}
\end{center}
\end{figure}

In reality, instead of having a continuous dipole field the storage ring will be built with several dipole magnets, powered in series to define the geometry of the ring.
For a single magnet, the trajectory of a particle is shown schematically  in Fig.~\ref{TSR_dipole_field}.
In the free space outside the dipole magnet, the particle trajectory follows a straight line. As soon as the particle enters the magnet, it is bent onto a circular path until it leaves the magnet at the other side.

The overall effect of the main bending (or `dipole') magnets in the ring is to define a more or less circular path, which we call the `design orbit'. By definition, this design orbit has to be a closed loop,  and so  the main dipole magnets in the ring have to define a full bending angle of exactly $2 \pi$. If $\alpha$ denotes the bending angle of a single magnet, then
\begin{equation}
\alpha=\frac{\rmd s}{\rho}= \frac{B \, \rmd s}{B \cdot \rho}.
\end{equation}
We  therefore require that
\begin{equation}
\frac{\int B \, \rmd s}{B \cdot \rho} = 2 \pi.
\label{eq_beam_rigidity}
\end{equation}
Thus, a storage ring or synchrotron is not a `ring' in the true sense of the word but more a polygon, where `poly' means the discrete number of dipole magnets installed in the `ring'.

\begin{figure}[h]
\begin{center}
\includegraphics[width=0.5\columnwidth] {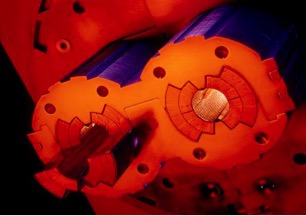}
\caption{Superconducting dipole magnet in the LHC storage ring}
\label{LHC_dip}
\end{center}
\end{figure}

In the case of the LHC, the dipole field has been pushed to the highest achievable values:
1232 superconducting dipole magnets, each 15~m long, define the geometry of the ring and, via  Eq.~(\ref{eq_beam_rigidity}),  the maximum momentum for the stored proton beam.
Using the equation given above, for a maximum momentum $p=7$~TeV/$c$, we obtain a required magnetic field of
\begin{equation}
B=\frac {2 \pi \cdot 7000 \cdot 10^{9}~{\rm eV}}{1232 \cdot 15~{\rm m} \cdot 2.99792 \cdot 10^{8}~{\rm m}\,{\rm s}^{-1}},
\end{equation}
or
\begin{equation}
B=8.33~{\rm T},
\end{equation}
 to bend the beams.
For convenience, we have expressed the particle momentum in units of GeV/$c$ here.
Figure~\ref{LHC_dip} shows a photograph of one of the LHC dipole magnets, built with superconducting NbTi filaments, which are operated at a temperature $T=1.9$~K.

\subsection{Focusing properties}
In addition to the main bending magnets that guide the beam onto a closed orbit, focusing fields are needed to keep the particles close together. In modern storage rings and light sources, we have to keep more than $10^{12}$ particles in the machine, distributed over a number of bunches, and these particles have to be focused to keep their trajectories close to the design orbit. Furthermore, these particles are stored in the machine for many hours, and a carefully designed focusing structure is needed to maintain the necessary beam size at different locations in the ring and guarantee stability of the transverse motion.

\begin{figure}[h]
\begin{center}
\includegraphics[width=0.5\columnwidth] {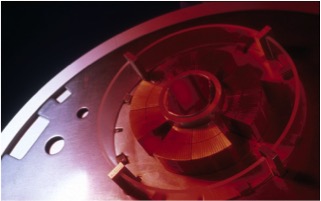}
\caption{Superconducting quadrupole magnet in the LHC storage ring}
\label{LHC_quad}
\end{center}
\end{figure}

Following classical mechanics, linear restoring forces are used, just as in the case of a harmonic pendulum. Quadrupole magnets provide the corresponding  field property:
they create a magnetic field that depends linearly on the amplitude of the particle, i.e., the distance of the particle from the design orbit:
\begin{equation}
B_{x}=g\cdot y , \qquad
B_{y}=g \cdot x .
\end{equation}
The constant $g$ is called the gradient of the magnetic field and characterizes the focusing strength of the quadrupole lens in both transverse planes. For convenience, it is normalized (like the dipole field) to the particle momentum. This normalized gradient is denoted by $k$ and defined as
\begin{equation}
k=\frac{g}{p/e}=\frac{g}{B \rho}.
\label{kdefinition}
\end{equation}
The technical layout of such a quadrupole is depicted in Fig.~\ref{LHC_quad}. As in the case of the  dipoles, the LHC quadrupole magnets were built using superconducting technology.

\begin{figure}[h]
\begin{center}
\includegraphics[width=0.6\columnwidth]{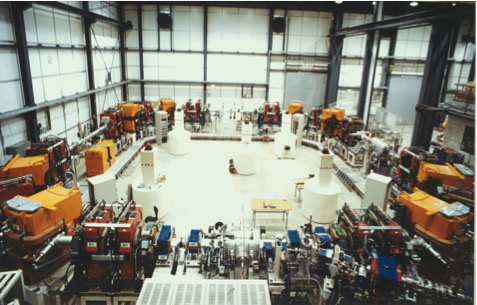}
\caption{The TSR storage ring, Heidelberg, as a typical example of a separate-function strong-focusing storage ring \cite{TSR}.}
\label{TSR_photo}
\end{center}
\end{figure}

Now that we have defined the two basic building blocks of a storage ring, we need to arrange them in a so-called magnet lattice and optimize the field strengths in such a way as to obtain the required beam parameters. An example of how such a magnet lattice looks like is given in Fig.~\ref{TSR_photo}. This photograph shows the dipole (orange) and quadrupole (red) magnets in the TSR storage ring in Heidelberg \cite{TSR}.  Eight dipoles are used to bend the beam into a `circle', and the quadrupole lenses between them provide the focusing to keep the particles within the aperture limits of the vacuum chamber.

A general design principle of modern synchrotrons and storage rings should be pointed out here.
In general, these machines are built following a so-called separate-function scheme: every magnet is designed and optimized for a certain task, such as bending, focusing, or chromatic correction. We separate the magnets in the design according to the job they are supposed to do; only in rare cases a combined-function scheme is chosen nowadays, where different magnet properties are combined in one piece of hardware. To express this principle mathematically, we use the general Taylor expansion of the normalized magnetic field,
\begin{equation}
\frac{B(x)}{p/e}=\frac{1}{\rho}+ k \cdot x + \frac{1}{2 !}mx^{2} + \frac{1}{3 !}nx^{3}+ \cdots.
\end{equation}
Following the arguments above, for the moment we take into account only constant (dipole) or linear  (quadrupole) terms.
The higher-order contributions to the field will be treated later as (hopefully) small perturbations.

\begin{figure}[h]
\begin{center}
\includegraphics[width=0.5\columnwidth]{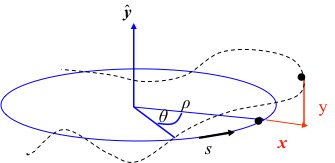}
\caption{Coordinate system used in particle beam dynamics; the longitudinal coordinate $s$ moves  around the ring with the particle considered.%
}
\label{Coordsystem}
\end{center}
\end{figure}

Under these assumptions, we can derive --- in a linear approximation --- the equation of motion of the transverse particle movement.
To derive the equation of motion, we start with a general expression for the radial acceleration as known from classical mechanics (see, e.g., \cite{Goldstein}):
\begin{equation}
a_{r} = \frac{\rmd^{2} \rho}{\rmd t^{2}} - \rho \left(\frac{\rmd \theta}{\rmd t} \right)^{2}.
\end{equation}
The first term refers to an explicit change in the bending radius, and the second to the centrifugal acceleration. Referring to our coordinate system, and replacing the ideal radius $\rho$ by $\rho + x$ for the general case  (Fig.~\ref{Coordsystem}), we obtain for the balance between the radial force and the counter-acting Lorentz force the relation
\begin{equation}
F=m \frac{\rmd^{2}}{\rmd t^{2}}(x+\rho)- \frac{mv^{2}}{x+\rho}=evB.
\end{equation}

On the right-hand side of the equation, we take only linear terms of the magnetic field into account,	
\begin{equation}
B_{y}=B_{0}+x \frac{\rmd B_{y}} {\rmd x},
\end{equation}
and for convenience we replace the independent variable $t$ by the coordinate  $s$,
\begin{equation}
x'=\frac{\rmd x} {\rmd s}=\frac{\rmd x} {\rmd t} \frac{\rmd t} {\rmd s} ,
\end{equation}
to obtain an expression for the particle trajectories under the influence of the focusing properties of the quadrupole and dipole fields in the ring, described by a differential equation. This equation is derived in its full beauty elsewhere \cite{Goldstein}, so we shall just state it here:
\begin{equation}
x''+x \cdot \left(\frac {1}{\rho^{2}}+k\right)=0,
\label{eom}
\end{equation}
where $k$ is the normalized gradient introduced above and the $1/ \rho^{2}$ term represents the so-called weak focusing, which is a property of the bending magnets.

The particles will now follow the `circular' path defined by the dipole fields, and in addition will undergo harmonic oscillations in both transverse planes. The situation is shown schematically in Fig.~\ref{Coordsystem}. An ideal particle will follow the design orbit represented by the circle in the diagram. Any other particle will perform transverse oscillations under the influence of the external focusing fields,  and the amplitude of these oscillations will ultimately define the beam size.

\begin{figure}[h]
\begin{center}
\includegraphics[width=0.4\columnwidth]{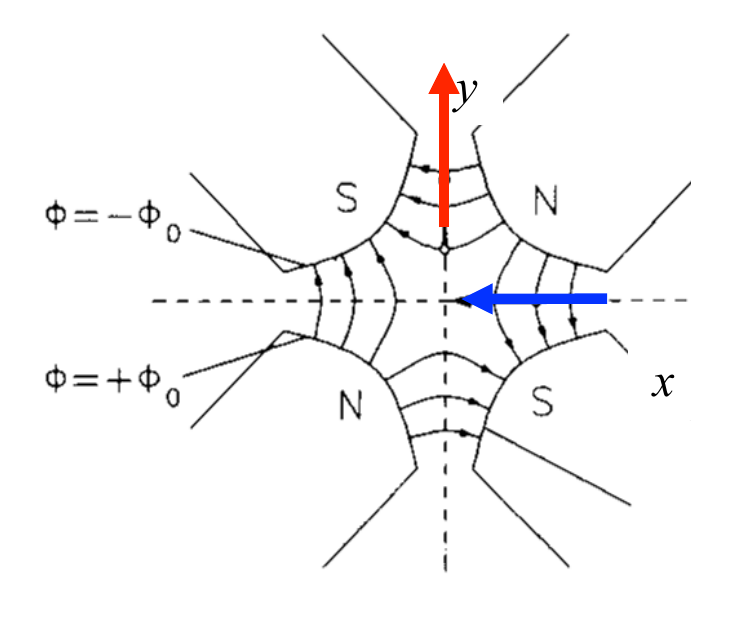}
\caption{Field configuration in a quadrupole magnet and the direction of the focusing and defocusing forces in the horizontal and vertical planes.%
}
\label{Quad_field}
\end{center}
\end{figure}

Unlike the case of a classical harmonic oscillator, however, the equations of motion in the horizontal and vertical planes differ somewhat.
Assuming a horizontal focusing magnet, the equation of motion is as shown in Eq. (\ref{eom}).
In the vertical plane, on the other hand, because of the orientation of the field lines and thus by Maxwell's equations, the forces instead have a defocusing effect. Also, the weak focusing term disappears in general:
\begin{equation}
y''- y \cdot k=0.
\end{equation}
The principal problem arising from the different directions of the Lorentz force in the two transverse planes of a quadrupole field  is sketched  in Fig.~\ref{Quad_field}. So, we have to explicitly introduce quadrupole lenses that focus the beam in the horizontal and vertical directions in some alternating order, and it is the task of the machine designer to find an adequate solution to this problem and to define a magnet pattern that will provide an overall focusing effect in both transverse planes.

Following closely the example of the classical harmonic oscillator,
we can write down the solutions of the above  equations of motion. For simplicity, we focus on the  horizontal plane; a `focusing' magnet is therefore focusing in this horizontal plane and at the same time defocusing in the vertical plane.
Starting with the initial conditions for the particle amplitude $x_{0}$ and angle $x'_{0}$ in front of the magnet element, we obtain the following relations for the trajectory inside the magnet:
\begin{align}
x(s)& = x_{0}\cdot \cos\bigl(\sqrt{{|K|}}\,s\bigr) +x'_{0} \cdot \frac{1}{\sqrt{{|K|}}}\sin\bigl(\sqrt{{|K|}}\,s\bigr), \\
x'(s)&=-x_{0}\cdot \sqrt{{|K|}} \sin\bigl(\sqrt{{|K|}}\,s\bigr)+x'_{0}\cdot \cos\bigl(\sqrt{{|K|}}\,s\bigr).
\end{align}
Here the parameter $K$ combines the quadrupole gradient and the weak focusing effect:
$K = k - 1/\rho^{2}$.
Usually, these two equations are combined into a more elegant and convenient matrix form,
\begin{equation}
\begin{pmatrix}
x \\
x'
\end{pmatrix}_{s}
=\mathbf{M}_{\rm foc}
\begin{pmatrix}
x \\
x'
\end{pmatrix}_{0},
\end{equation}
where the matrix $\mathbf{M}_{\rm foc}$ contains all the relevant information about the magnet element:

\begin{equation}
\mathbf{M}_{\rm foc} =
\begin{pmatrix}
\cos(\sqrt{{|K|}}\,s)              &     \frac{1}{\sqrt{{|K|}}} \sin(\sqrt{{|K|}}\,s)  \\
- \sqrt{{|K|}}\sin(\sqrt{{|K|}}\,s)   &   \cos(\sqrt{{|K|}}\,s)
\end{pmatrix}.
\end{equation}

\begin{figure}[h]
\begin{center}
\includegraphics[width=0.6\columnwidth]{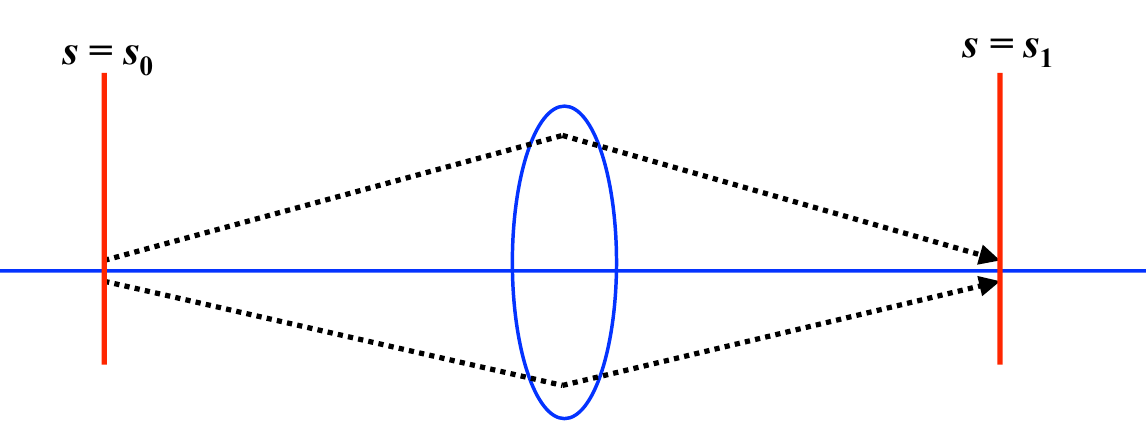}
\caption{Schematic illustration of the principle of the effect of a focusing quadrupole magnet}
\label{Matrix_foc}
\end{center}
\end{figure}

The situation is visualized schematically in Fig.~\ref{Matrix_foc}.

\begin{figure}[h]
\begin{center}
\includegraphics[width=0.6\columnwidth]{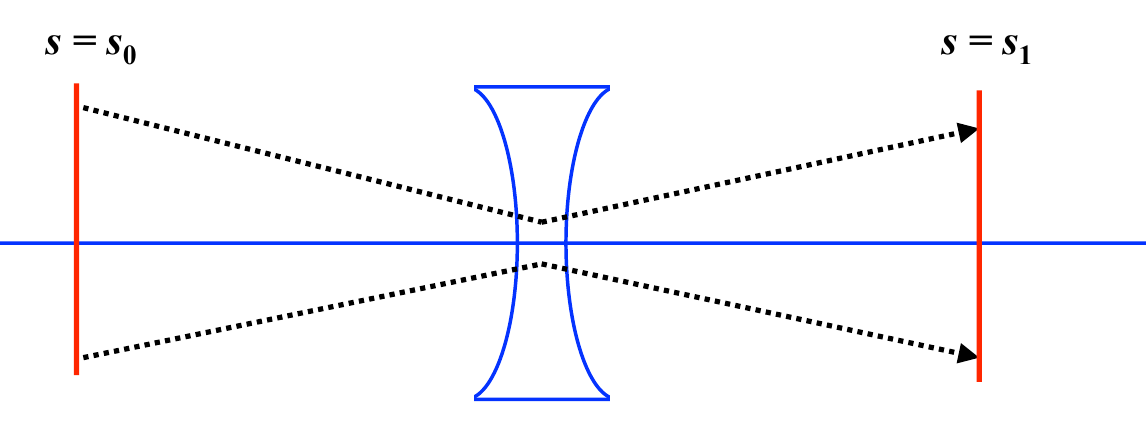}
\caption{Schematic illustration of the principle of the effect of a defocusing quadrupole magnet}
\label{Matrix_defoc}
\end{center}
\end{figure}

In the case of a defocusing magnet, we obtain  analogously that
\begin{equation}
\begin{pmatrix}
x \\
x'
\end{pmatrix}_{s}
=\mathbf{M}_{\rm defoc}
\begin{pmatrix}
x \\
x'
\end{pmatrix}_{0},
\end{equation}
with
\begin{equation}
\mathbf{M}_{\rm defoc}=
\begin{pmatrix}
\cosh(\sqrt{{|K|}}\,s)              &     \frac{1}{\sqrt{{|K|}}} \sinh(\sqrt{{|K|}}\,s)  \\
\sqrt{{|K|}}\sinh(\sqrt{{|K|}}\,s)   &   \cosh(\sqrt{{|K|}}\,s)
\end{pmatrix};
\end{equation}
see Fig.~\ref{Matrix_defoc}.

For completeness, we also include the case of a field-free drift. With $K=0$, we obtain
\begin{equation}
\mathbf{M}_{\rm drift}=
\begin{pmatrix}
1          &     s  \\
0          &     1
\end{pmatrix}.
\end{equation}

\begin{figure}[h]
\begin{center}
\includegraphics[width=0.45\columnwidth]{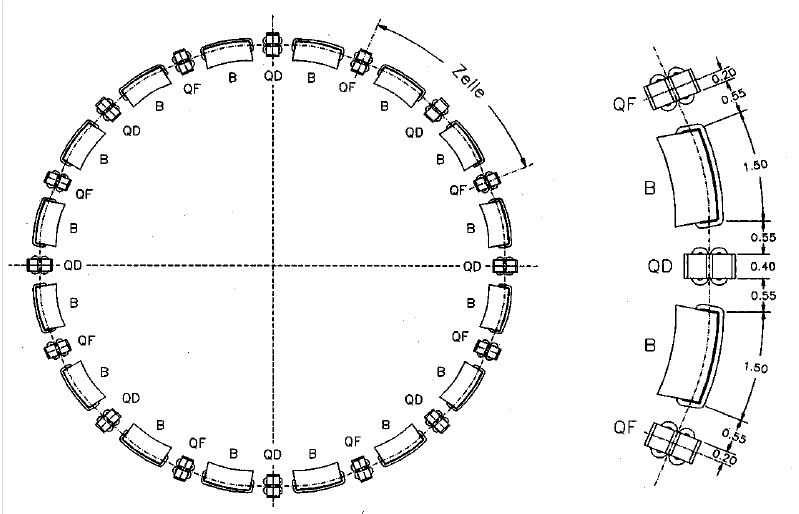}
\caption{A simple periodic chain of bending magnets and focusing/defocusing quadrupoles forming the basic structure of a storage ring ~\cite{Wille}.}
\label{Willering}
\end{center}
\end{figure}

This matrix formalism allows us  to combine the elements of a storage ring in an elegant way, and so it is straightforward to calculate particle trajectories. As an example, we consider the simple case of an alternating focusing and defocusing lattice, a so-called FODO lattice \cite{Wille}; see Fig.~\ref{Willering}.

\begin{figure}[h]
\begin{center}
\includegraphics[width=0.8\columnwidth]{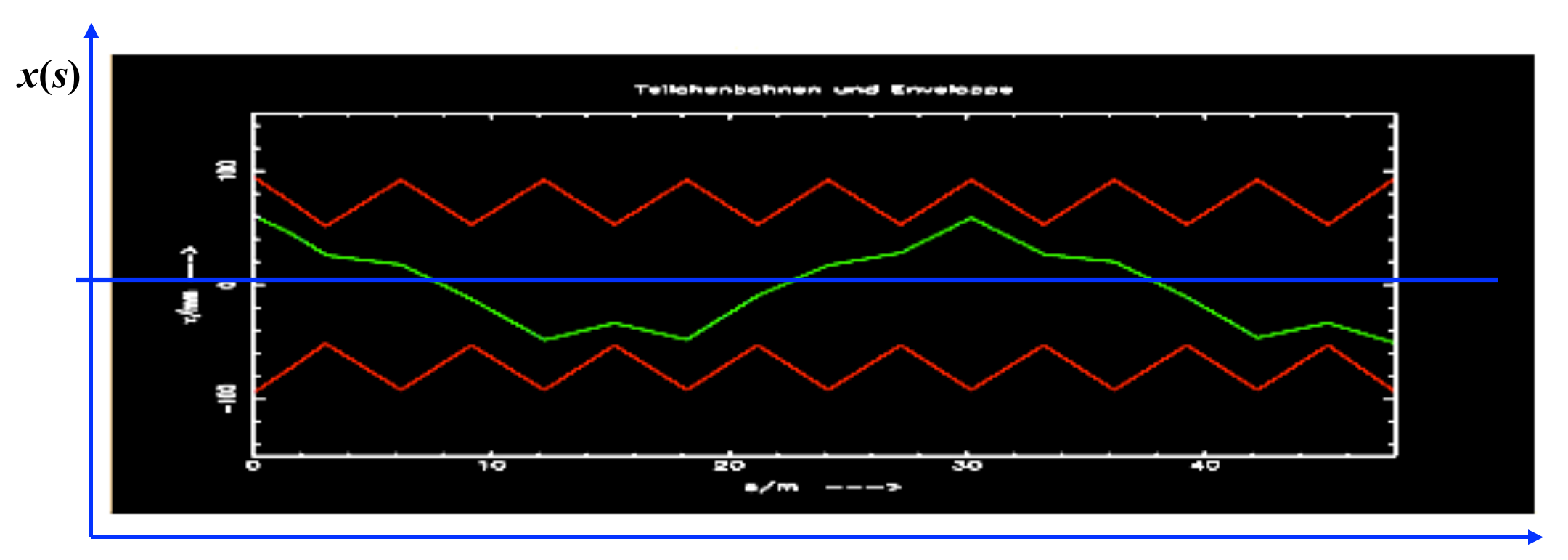}
\caption{Calculated particle trajectory in a simple storage ring}
\label{track_1}
\end{center}
\end{figure}

As we know the properties of each and every element in the accelerator, we can construct the corresponding matrices and calculate step by step the amplitude and angle of a single-particle trajectory around the ring.
Even more conveniently, we can multiply out the different matrices and, given initial conditions $x_{0}$ and $x'_{0}$, obtain directly the trajectory at any location in the ring:
\begin{equation}
\mathbf{M}_{\rm total} =  \mathbf{M}_{\rm foc} \cdot \mathbf{M}_{\rm drift}\cdot \mathbf{M}_{\rm dipole}\cdot \mathbf{M}_{\rm drift} \cdot \mathbf{M}_{\rm defoc}\cdots.
\end{equation}
The trajectory thus obtained is shown schematically in Fig.~\ref{track_1}.

We emphasize the following facts  in this context.
\begin{itemize}
\item At each moment, which means inside each lattice element, the trajectory is a part of a harmonic oscillation.
\item However, because of the different restoring or defocusing forces, the solution will look different at each location.
\item In the linear approximation that we have made use of in this context, all particles experience the same external fields, and their trajectories will differ only because of their different initial conditions.
\item There seems to be an overall oscillation in both transverse planes while the particle is travelling around the ring. Its amplitude stays well within the boundaries set by the vacuum chamber, and its frequency in the example of Fig.~\ref{track_1} is roughly 1.4 transverse oscillations per revolution, which corresponds to the eigenfrequency of the particle under the influence of the external fields.
\end{itemize}

\begin{figure}[h]
\begin{center}
\includegraphics[width=0.65\columnwidth]{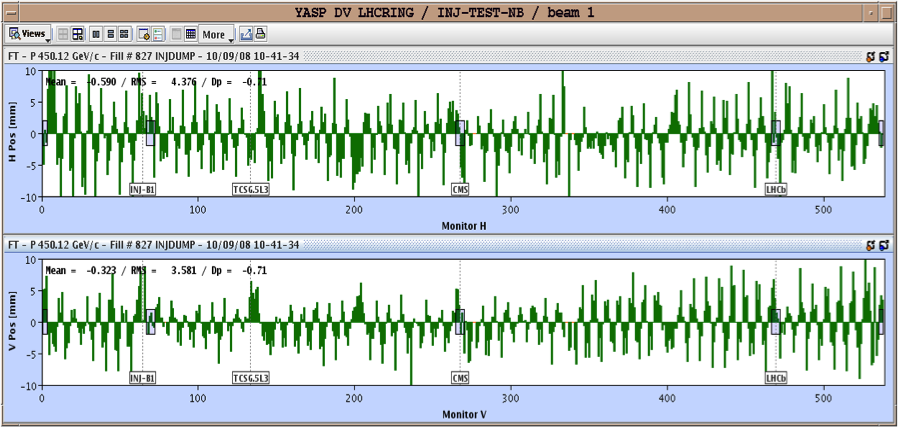}
\caption{Measured orbit in LHC during the commissioning of the machine}
\label{LHC_orbit}
\end{center}
\end{figure}

Coming closer to a real, existing machine, we show in Fig.~\ref{LHC_orbit} an orbit measured during one of the first injections into the LHC storage ring. The horizontal oscillations are plotted in the upper half of the figure and the vertical oscillations in the lower half, on a scale of $\pm 10$~mm. Each histogram bar indicates the value recorded by a beam position monitor at a certain location in the ring,  and the orbit oscillations are clearly visible. By counting (or, better, fitting) the number of oscillations in both transverse planes, we obtain values of
\begin{equation}
Q_{x}=64.31, \qquad
Q_{y}=59.32.
\end{equation}
These values, which describe the eigenfrequencies of the particles, are called the horizontal and vertical \emph{tunes}, respectively. Knowing the revolution frequency,  we can easily calculate the transverse oscillation frequencies, which for this type of machine usually lie in the range of some hundreds of kilohertz.

\begin{figure}[h]
\begin{center}
\includegraphics[width=0.65\columnwidth]{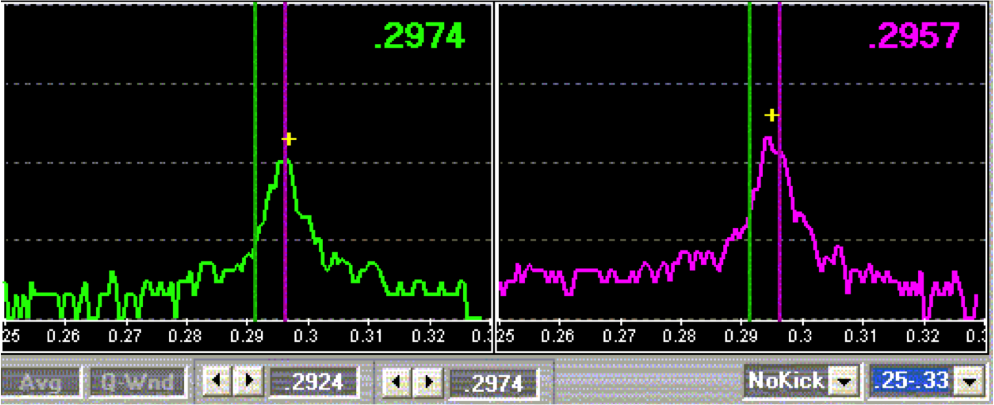}
\caption{Tune signal of a proton storage ring (HERA-p)}
\label{HERA_tune}
\end{center}
\end{figure}

As the tune characterizes the particle oscillations under the influence of all external fields, it is one of the most important parameters of a storage ring. Therefore it is usually  displayed and controlled at all times by the control system of such a machine. As an example, Fig.~\ref{HERA_tune} shows  the tune diagram of the HERA proton ring \cite{HERA}; this was obtained via a Fourier analysis of a spectrum measured from the signal of the complete particle ensemble. The peaks indicate the two tunes in the horizontal and vertical planes of the machine, and in a sufficiently linear machine a fairly  narrow spectrum is obtained.

\begin{figure}[h]
\begin{center}
\includegraphics[width=0.65\columnwidth]{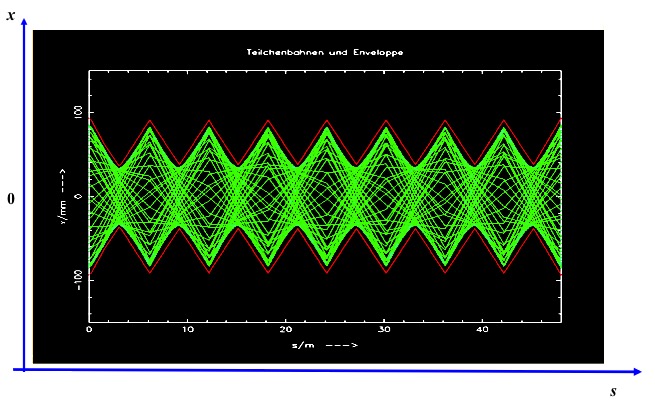}
\caption{Many single-particle trajectories together form a pattern that corresponds to the beam size in the ring.}
\label{track_n}
\end{center}
\end{figure}

Briefly referring back to Fig.~\ref{track_1}, the question is what the trajectory of the particle will look  like in the second turn, or the third, or after an arbitrary number of turns. Now, as we are dealing with a circular machine, the amplitude and angle $x$ and $x'$ at the end of the first turn will be the initial conditions for the second turn, and so on. After many turns, the overlapping trajectories begin to form a pattern, such as that in Fig.~\ref{track_n},  which indeed looks like a beam that here and there has a larger and a smaller size but still remains well defined in its amplitude by the external focusing forces.

\section{The Twiss parameters $\alpha$, $\beta$, and $\gamma$}

As explained above, repeating the calculations that lead to the orbit of the first turn will result in a large number of single-particle trajectories that overlap in some way and form the beam envelope. Figure~\ref{track_n}  shows the result for 50 turns. Clearly, as soon as we are talking about many turns or many particles, the use of the single-trajectory approach is quite limited and we need a description of the beam as an ensemble of many particles.
Fortunately, in the case of periodic conditions in the accelerator, there is another way to describe the particle trajectories and, in many cases, it is more convenient than the above-mentioned formalism. It is important to note that, in a circular accelerator, the focusing elements are necessarily periodic in the orbit coordinate $s$ after one revolution. Furthermore, storage ring lattices have an internal periodicity in most cases: they are often constructed, at least partly, from sequences in which identical magnetic structures, the lattice cells, are repeated several times in the ring and lead to periodically repeated focusing properties.
In this case, the equation of motion can now be written in a slightly different form:
\begin{equation}
x''(s) - k(s) \cdot x(s) =0,
\end{equation}
where, for simplicity, we refer to a pure quadrupole magnet and so the $1/\rho^{2}$ term does not appear. The main issue, however, is that unlike the treatment above, the focusing parameters (or restoring forces) are no longer constant but are functions of the coordinate $s$. However, they are periodic in the sense that, at least after one full turn, they repeat themselves, i.e., $k(s + L) = k(s)$, leading to the so-called Hill differential equation. Following Floquet's theorem  \cite{floquet},  the solution of this equation can be written in its general form as
\begin{equation}
x(s)=\sqrt{\varepsilon}\sqrt{\beta(s)} \cos(\psi(s) + \phi),
\label{epsbeta}
\end{equation}
where $\psi$ is the phase of the oscillation, $\phi$ is its initial condition, and $\varepsilon$ is a characteristic parameter of a single particle or, if we are considering a complete beam, of the ensemble of particles.
Taking the derivative with respect to $s$, we get
\begin{equation}
x'(s)= - \frac{\sqrt{\varepsilon}}{\sqrt{\beta(s)}} {\sin(\psi(s) - \phi) + \cos(\psi(s) + \phi)} .
\label{epsbetastr}
\end{equation}

The position and angle of the transverse oscillation of a particle at a point $s$ are given by the value of a special amplitude function, the $\beta$-function, at that location, and  $\varepsilon$ and $\phi$ are constants of any particular trajectory. The $\beta$-function depends in a rather complicated manner on the overall focusing properties of the storage ring. It cannot be calculated directly by an analytical approach, but instead has to be either determined numerically or deduced from properties of the single-element matrices described above (see, e.g., \cite{bjh_erice}). In any case, like the lattice itself, it has to fulfil the periodicity condition
\begin{equation}
\beta(s+L)= \beta(s) .
\label{betaperiod}
\end{equation}
Inserting the solution (\ref{epsbeta}) into the Hill equation and rearranging slightly, we get
\begin{equation}
\psi(s) = \int_0^s \frac{\rmd s}{\beta(s)} ,
\label{psi}
\end{equation}
which describes the phase advance of the oscillation. It should be emphasized that $\psi$ depends on the amplitude of oscillation of the particle. At locations where $\beta$ reaches large values, i.e., the beam has a large transverse dimension, the corresponding phase advance is small; conversely, at locations where we create a small $\beta$ in the lattice, we obtain a large phase advance. In the context of Fig.~\ref{track_1},  we introduced the tune as the number of oscillations per turn, which is nothing other than the overall phase advance of the transverse oscillation per revolution in units of 2$\pi$. So, by integrating Eq.~(\ref{psi})  around the ring, we get the expression
\begin{equation}
Q = \frac{1}{2\pi}  \oint \frac{\rmd s}{\beta(s)} .
\label{tune}
\end{equation}

\begin{figure}[h]
\begin{center}
\includegraphics[width=0.48\columnwidth]{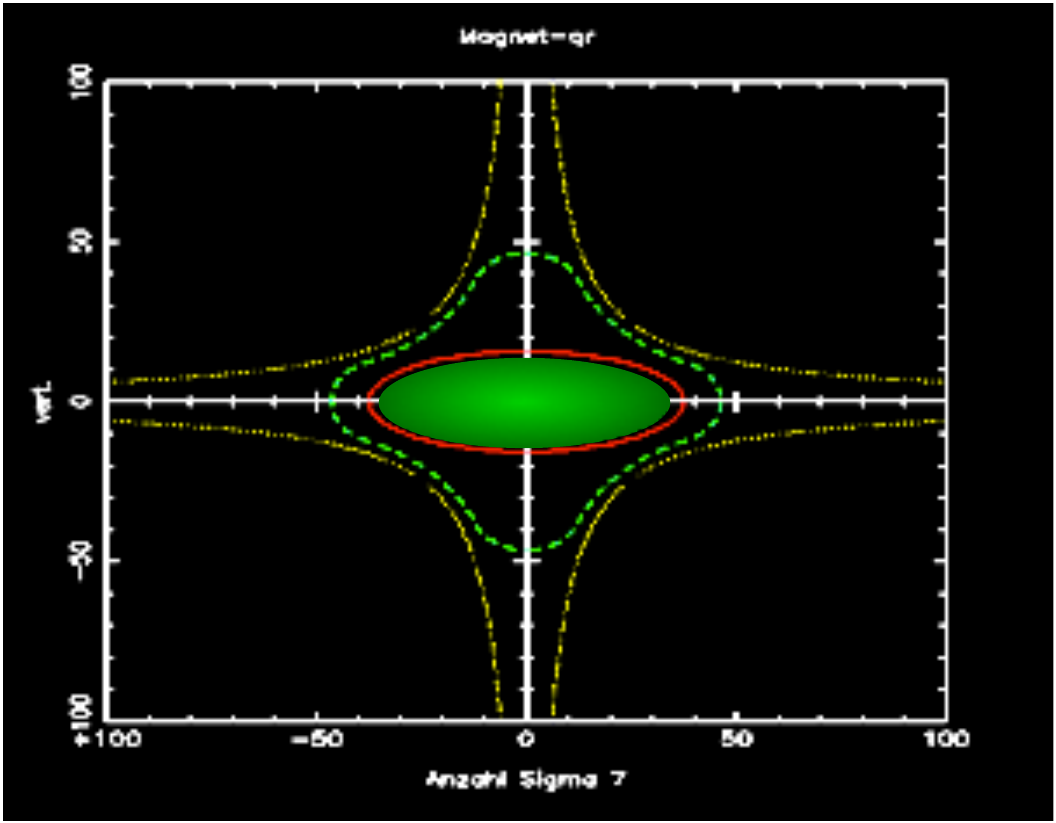}
\caption{Transverse beam shape inside a quadrupole magnet}
\label{envelope2}
\end{center}
\end{figure}

The practical significance of the $\beta$-function is shown in Figs.~\ref{track_n} and \ref{envelope2}. Whereas in Fig.~\ref{track_n} the single-particle trajectories are plotted turn by turn, Fig.~\ref{envelope2} shows schematically a section through the transverse shape of the beam and indicates the beam size inside the vacuum chamber. The hyperbolic profile of the pole shoes of the quadrupole lens is sketched as a yellow dashed line, and the envelope of the overlapping trajectories,  given by $\hat{x}=\sqrt{\varepsilon \beta(s) }$, is marked in red and used to define the beam size in the sense of a Gaussian density distribution.

\subsection{$\beta, \varepsilon$, and the phase space ellipses}

Although the $\beta$-function is a somewhat abstract parameter that results from all focusing and defocusing elements in the ring, the integration constant $\varepsilon$ has a well-defined physical interpretation. Given the solution of Hill's equation
\begin{equation}
x(s)=\sqrt{\varepsilon}\sqrt{\beta(s)}\cos(\psi(s) + \phi)
\label{epsbeta2}
\end{equation}
and its derivative
\begin{equation}
x'(s)= - \frac{\sqrt{\varepsilon}}{\sqrt{\beta(s)}} {\sin(\psi(s) - \phi) + \cos(\psi(s) + \phi)},
\label{epsbetastr1}
\end{equation}
we can transform Eq. (\ref{epsbeta2}) to
\begin{equation}
\cos(\psi(s)) = \frac{x(s)}{\sqrt{\varepsilon \beta(s)}}
\label{epsbetastr2}
\end{equation}
and insert the expression into Eq.~(\ref{epsbetastr1}) to get an expression for the integration constant $\varepsilon$:
\begin{equation}
\varepsilon = \gamma(s) x^{2}(s) + 2 \alpha x(s) x'(s) +\beta(s)x'^{2}(s) .
\label{ellipseparametric}
\end{equation}
Here we have followed the usual convention in the literature and introduced the two parameters
  \begin{equation}
\alpha(s) = - \frac{1}{2}\beta'(s)
\end{equation}
and
\begin{equation}
\gamma(s)=\frac{1+\alpha^{2}(s)}{\beta(s)} .
\end{equation}

We obtain for $\varepsilon$ a parametric representation of an ellipse in the $(x, x')$ `phase space', which, according to Liouville's theorem, is a constant of motion, as long as only conservative forces are considered. The mathematical integration constant thus gains physical meaning.
In fact, $\varepsilon$ describes the space occupied by the particle in the transverse $(x, x')$ phase space (simplified here to a two-dimensional space). More specifically, the area  in the $(x, x')$ space that is covered by the particle is given by
\begin{equation}
    A=\pi \cdot \varepsilon,
\end{equation}
and, as long as we consider only conservative forces acting on the particle, this area is constant according to Liouville's theorem. Here we take these facts as given, but we should point out that, as a direct consequence, the so-called emittance $\varepsilon$ cannot be influenced by any external fields; it is a property of the beam, and we have to take it as given and handle it with care.

\begin{figure}[h]
\begin{center}
\includegraphics[width=0.5\columnwidth]{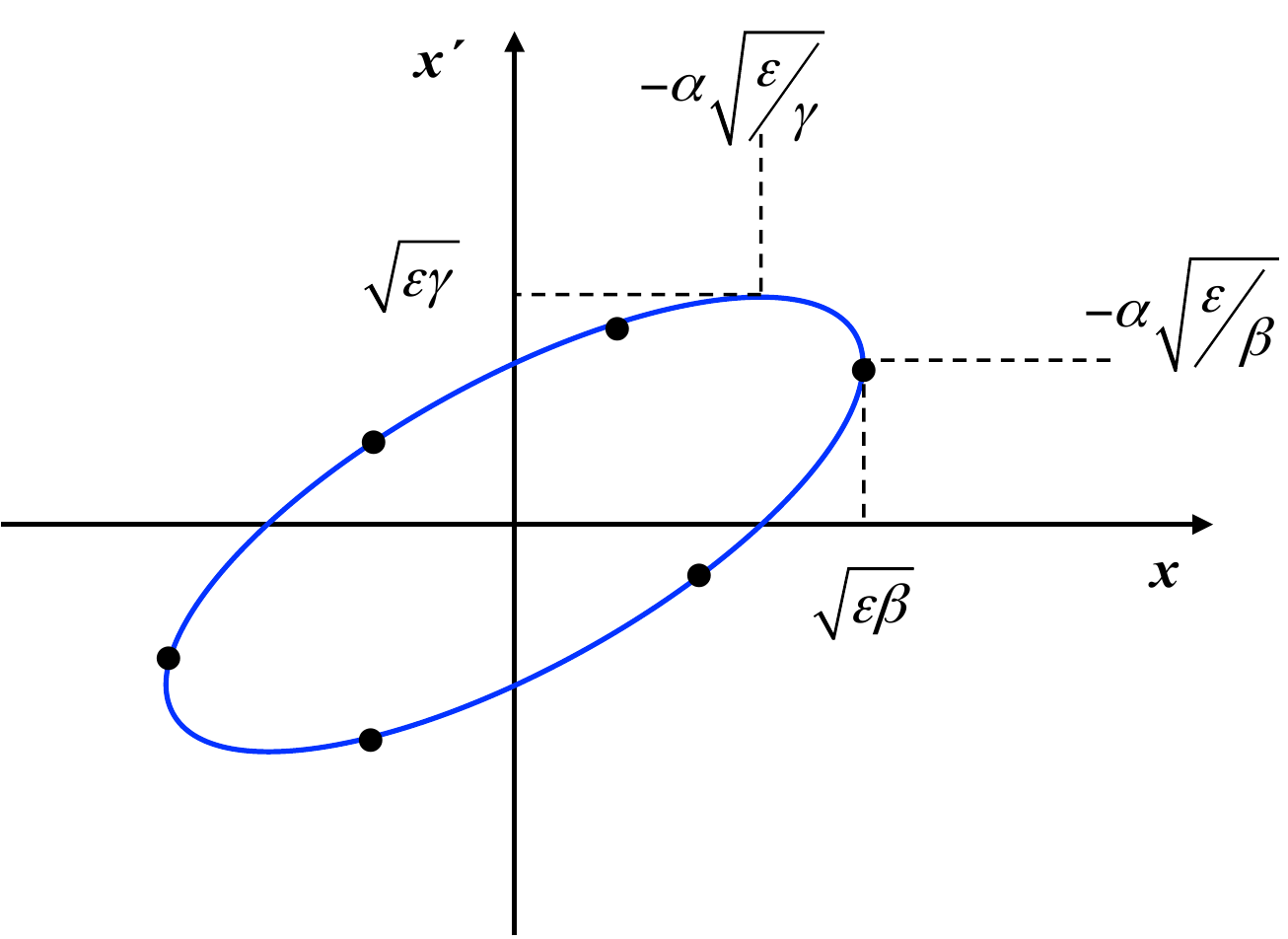}
\caption{Ellipse in $(x, x')$ phase space}
\label{Ellipse}
\end{center}
\end{figure}

To be more precise, and following the usual textbook treatment of accelerators, we can draw the ellipse of the particle's transverse motion in phase space; see, for example, Fig.~\ref{Ellipse}. Although the shape and orientation are determined by the optics function $\beta $ and its derivative, $\alpha=-\frac{1}{2}\beta'$,  and so change as a function of the position $s$, the area covered in phase space is constant.

In Fig.~\ref{Ellipse}, expressions for the dependence of the beam size and divergence and, as a consequence, the shape and orientation of the phase space ellipse are included. For the sake of simplicity, we shall not derive these expressions here but instead refer to \cite{bjh_erice}.

Referring again to the single-particle trajectory discussed above (see Fig.~\ref{track_1}), but now plotting for a given position $s$ in the ring the coordinates $x$ and $x'$ turn by turn, we obtain the phase space coordinates of the particle as shown in Fig.~\ref{Ellipse} (marked as dots in the figure). These coordinates follow the form of an ellipse, whose shape and orientation are defined by the optical parameters at the reference position $s$ in the ring. Each point in Fig.~\ref{Ellipse} represents the transverse coordinates for a certain turn at that position in the ring, and the particle performs from one turn to the next a number of revolutions in phase space that corresponds to its tune. We have already emphasized that, as long as only conservative forces are considered (i.e., no interaction between the particles in a bunch, no collisions with remaining gas molecules, no radiation effects, etc.), the size of the ellipse in $(x, x')$ space is constant and can be considered as a quality factor of a single particle. Large areas in $(x, x')$ space mean large amplitudes and angles of transverse particle motion, and we would consider this as meaning a low particle `quality'.

Let us now talk a little more about the beam as an ensemble of many (typically $10^{11}$) particles. Referring to Eq.~(\ref{epsbeta}), at a given position in the ring the beam size is defined by the emittance $\varepsilon$ and the function $\beta$. Thus, at a certain moment in time, the cosine term in Eq. (\ref{epsbeta}) will be equal to one and the amplitude of the trajectory will reach its maximum value. Now, if we consider a particle at one standard deviation (sigma) of the transverse density distribution, then by using the emittance of this reference particle we can calculate the size of the complete beam, in the sense that the complete area (within one sigma) of all particles in the $(x, x')$ phase space is surrounded (and thus defined) by our one-sigma candidate.
Thus the value $\sqrt{\varepsilon \cdot \beta(s)}$ defines the one-sigma beam size in the transverse plane.

\begin{figure}[h]
\begin{center}
\includegraphics[width=0.6\columnwidth]{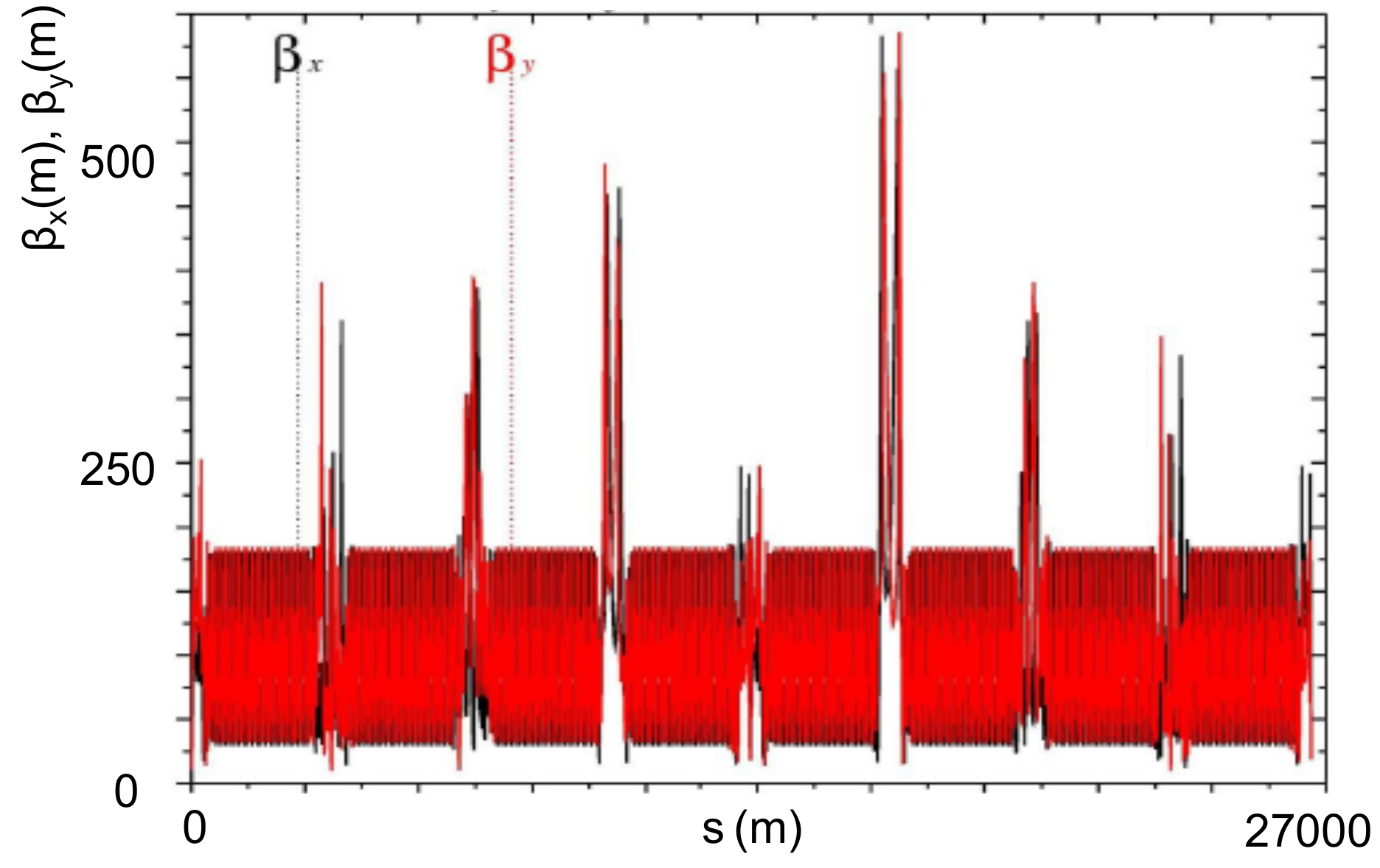}
\caption{LHC beam optics}
\label{lhc_optik}
\end{center}
\end{figure}
\begin{figure}[h]
\begin{center}
\includegraphics[width=0.6\columnwidth]{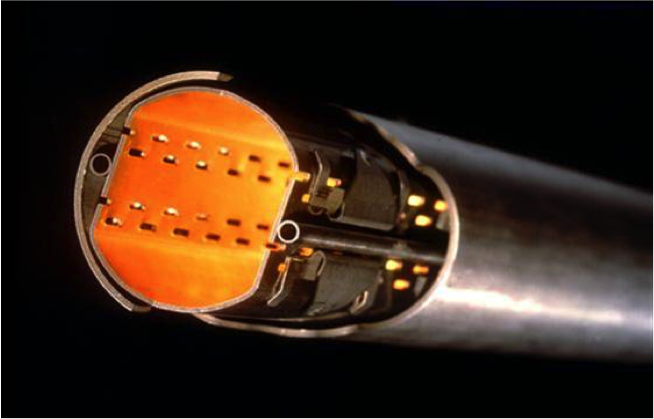}
\caption{The LHC vacuum chamber with a beam screen to shield the bore of the superconducting magnet from synchrotron radiation.}
\label{Beam_screen}
\end{center}
\end{figure}

As an example, we shall use the values for the LHC proton beam (Fig.~\ref{lhc_optik}). In the periodic pattern of the arc, the $\beta$-function is equal to 180~m and the emittance $\varepsilon$ at the flat-top energy is roughly $5 \times 10^{-10}$~rad\,m.
The resulting typical beam size is therefore 0.3~mm. Now, clearly, we would not design a vacuum aperture for the machine based on a one-sigma beam size; typically, an aperture requirement corresponding to 12$\sigma$ is a good rule to guarantee a sufficient beam lifetime, allowing for tolerances arising from magnet misalignments, optics errors, orbit fluctuations, and operational flexibility. In Fig.~\ref{Beam_screen}, part of the LHC vacuum chamber is shown, including the beam screen used to protect the cold bore from synchrotron radiation; this corresponds to a minimum beam size of $18\sigma$.

\section{Errors in field and gradient}

Up to now, we have treated the beam and the equation of motion as a monoenergetic problem. Unfortunately, in the case of a realistic beam, we have to deal with a considerable distribution of the particles with respect to energy or momentum. A typical value is
  \begin{equation}
\frac{\Delta p}{p}\approx 1.0 \cdot 10^{-3}.
\end{equation}
This momentum spread leads to several effects concerning the bending of the dipole magnets and the focusing strength of the quadrupoles. It turns out that the equation of motion, which has been a homogeneous differential equation until now, acquires a non-vanishing term on the right-hand side.

\subsection{Dispersive effects}

Replacing the ideal momentum $p$ in Eq.~(\ref{kdefinition}) by $p_0 +\Delta p$,  we obtain instead of  Eq.~(\ref{eom})
\begin{equation}
x''+x \cdot \left(\frac {1}{\rho^{2}}+k \right)=\frac{\Delta p}{p} \cdot \frac {1}{\rho} .
\end{equation}
The general solution of our now inhomogeneous differential equation is therefore the sum of the solution of the homogenous equation of motion and a particular solution of the inhomogeneous equation:
\begin{equation}
x(s) = x_\mathrm{h}(s) + x_\mathrm{i}(s) .
\end{equation}
Here $x_\mathrm{h} $ is the solution that we have discussed up to now and $x_\mathrm{i} $ is an additional contribution that still has to be determined. For convenience, we usually normalize this second term and define a special function, the so-called dispersion:
\begin{equation}
D(s) = \frac{x_\mathrm{i}(s)}{\Delta p / p} .
\end{equation}
This describes the dependence of the additional amplitude of the transverse oscillation on the momentum error of the particle. In other words, it fulfils the condition
\begin{equation}
x''_\mathrm{i}(s) + K(s) \cdot x_\mathrm{i}(s) = \frac {1}{\rho} \cdot \frac {\Delta p}{p} .
\end{equation}

The dispersion function is defined by the magnet lattice and is usually calculated by optics programs in the context of the calculation of the usual optical parameters; it is of equal importance. Analytically, it can be determined for single elements via the expression
\begin{equation}
D(s) = S(s)\cdot \int \frac {1}{\rho (\bar{s})} C(\bar{s}) \, \rmd \bar{s}  - C(s)\cdot \int \frac {1}{\rho (\bar{s})} S(\bar{s}) \, \rmd \bar{s} ,
\end{equation}
where $S(s)$ and $C(s)$ correspond to the sine-like and cosine-like elements of the single-element matrices or of the corresponding product matrix if there are several elements considered in the lattice.

\begin{figure}
\begin{center}
\includegraphics[width=0.65\columnwidth]{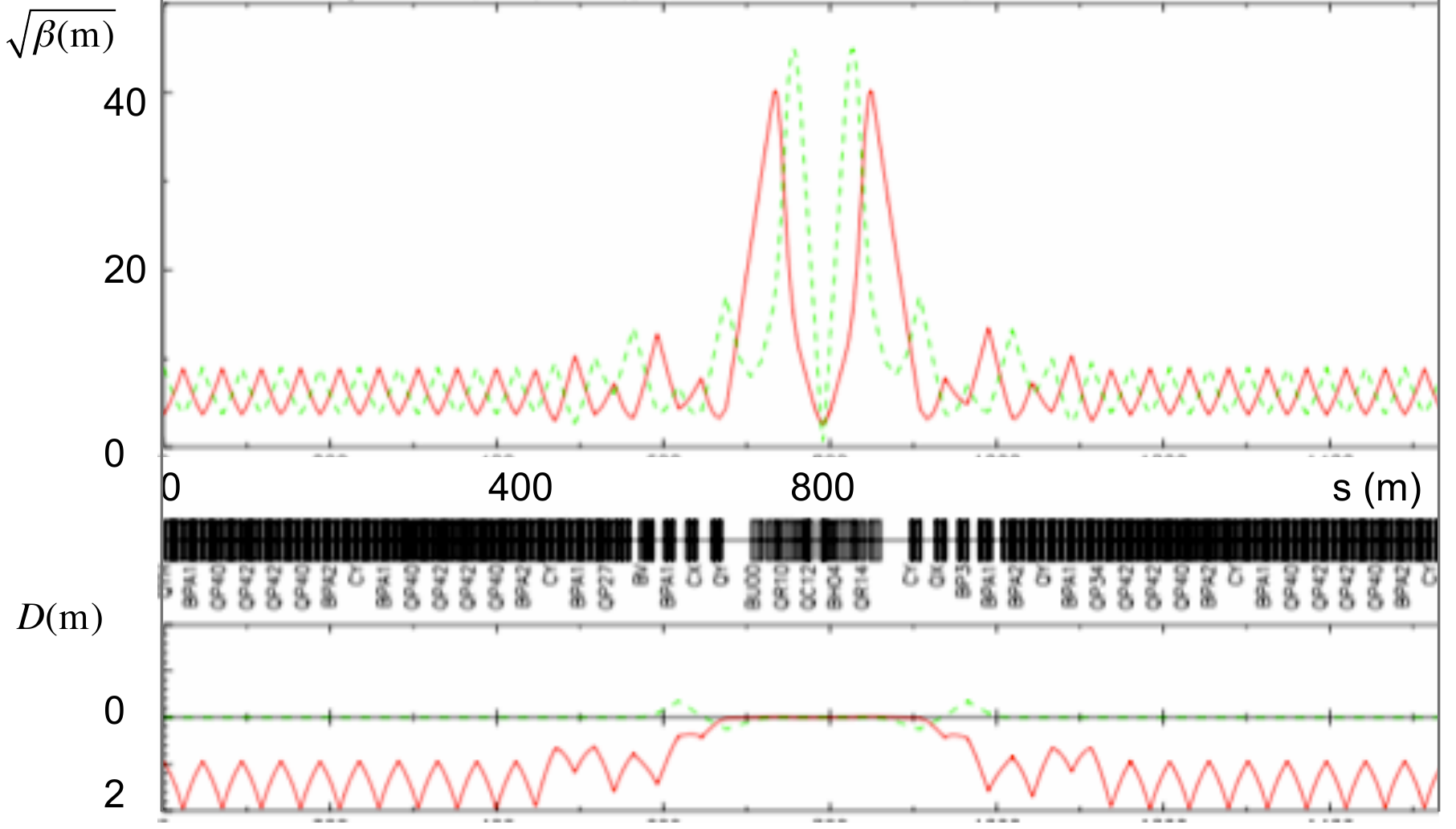}
\caption{$\beta$-function (upper part) and dispersion (lower part) of a typical high-energy collider ring}
\label{disp_plot}
\end{center}
\end{figure}

Although all this sounds somewhat theoretical, we would like to stress that typical values for the beam size and dispersive effect in the case of a high-energy storage ring are
\begin{equation}
x_{\beta} \approx \mbox{1--2~mm}, \quad D(s) \approx \mbox{1--2~m} .
\end{equation}
Thus, for a typical momentum spread of $\Delta p / p = 1 \cdot 10^{-3} $, we obtain an additional contribution to the beam size from the dispersion function that is of the same order as that from the betatron oscillations, $x_\beta$. An example of a high-energy beam optics system including the dispersion function is shown in Fig.~\ref{disp_plot}.
It should be pointed out that the dispersion describes the special orbit that an ideal particle would have in the absence of betatron oscillations ($x_\beta = x'_\beta = 0$) for a momentum deviation of $\Delta p/p = 1$. Nevertheless, it describes `just another particle orbit' and so it is subject to the focusing forces of the lattice elements, as seen
in the figure.

\subsection{Chromaticity}
\begin{figure}
\begin{center}
\includegraphics[width=0.65 \columnwidth]{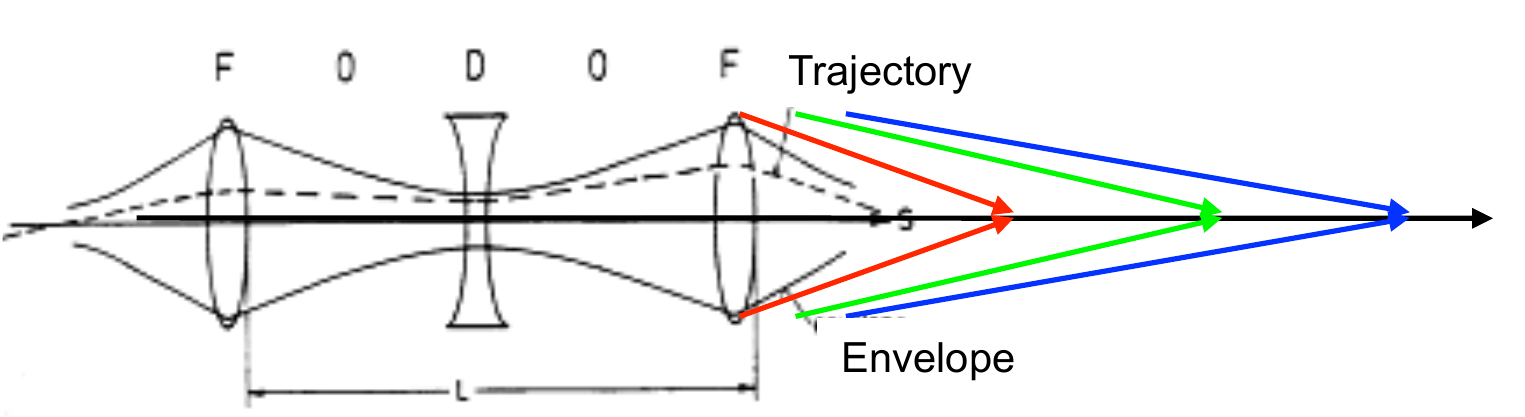}
\caption{Schematic view of the chromaticity effect in a quadrupole lens}
\label{Chroma_schematisch}
\end{center}
\end{figure}

Whereas dispersion is a problem that describes the non-ideal bending effect of dipoles in the case of a momentum error (or spread) in the particles, the careful reader will not be surprised to learn that a similar effect exists for the quadrupole focusing. We call this \emph{chromaticity}.
The chromaticity $Q'$ describes an optical error of a quadrupole lens in an accelerator: for a given
magnetic field, i.e., gradient of the quadrupole magnet, particles with a smaller momentum will feel a
stronger focusing force, and particles with a larger momentum will feel a
weaker force. The situation is shown schematically in Fig.~\ref{Chroma_schematisch}.
As a consequence, the tune of an individual particle will change, and the chromaticity $Q'$ relates the resulting tune shift to the relative momentum error of the particle. By definition, we write
\begin{equation}
\Delta Q = Q' \cdot \frac{\Delta p}{p} .
\end{equation}

$Q'$ is a consequence of the focusing properties of the quadrupole magnets and is thus given by the
characteristics of the lattice. For small momentum errors $\Delta p/p_0$, the focusing parameter $k$ can be written
as
\begin{equation}
k(p) =\frac{g}{p/e} = \frac{g e}{p_0+\Delta p} ,
\end{equation}
where $g$ denotes the gradient of the quadrupole lens, $p_0$ denotes the design momentum, and the term $\Delta p$ refers
to the momentum error. If $\Delta p$ is small, as we have assumed, we can write in a first-order approximation
\begin{equation}
k(p) \approx \frac{g e}{p_0}  \left(1 - \frac{\Delta p}{p_{0}} \right) .
\end{equation}
This describes a quadrupole error
\begin{equation}
\Delta k = -k_0 \cdot \frac{\Delta p}{p} .
\end{equation}
The negative sign indicates that a positive momentum deviation leads to a weaker focusing strength and, accordingly, to a negative tune shift:
\begin{equation}
\Delta Q = - \frac{1}{4 \pi} \int \Delta k \, \beta(s) \, \rmd s ,
\end{equation}
\begin{equation}
\Delta Q = - \frac{1}{4 \pi} \frac{\Delta p}{p}  \int  k_0 \beta(s) \, \rmd s .
\end{equation}
By definition, the chromaticity $Q'$ of a lattice is therefore given by
\begin{equation}
Q' = - \frac{1}{4 \pi}  \int  k(s)  \beta(s) \, \rmd s .
\label{qprime}
\end{equation}

\begin{figure}[h]
\begin{center}
\includegraphics[width=0.65\columnwidth]{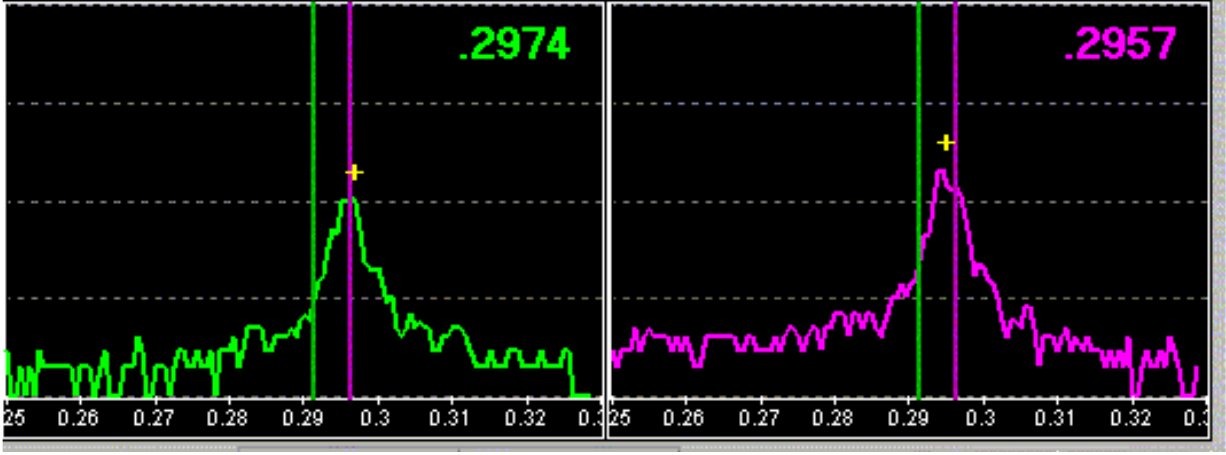}
\caption{Tune spectrum of a proton beam with a well-corrected chromaticity $Q' \approx 1$}
\label{chroma1}
\end{center}
\end{figure}

\begin{figure}[h]
\begin{center}
\includegraphics[width=0.65\columnwidth]{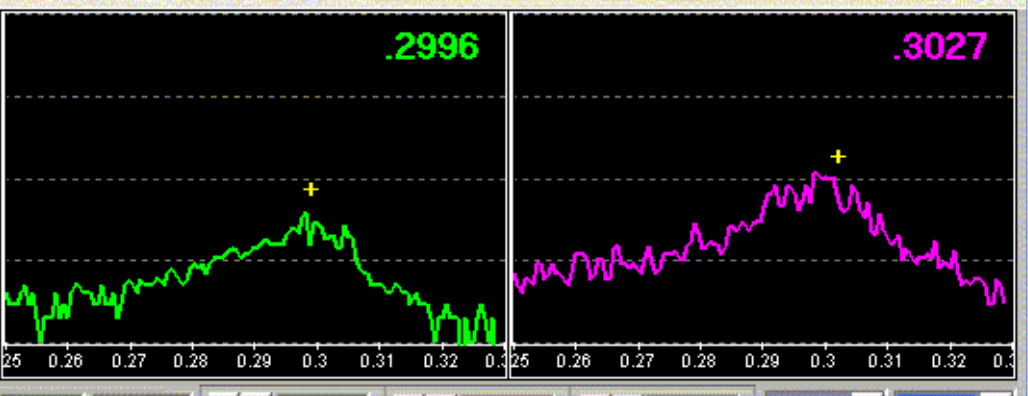}
\caption{Tune spectrum of a proton beam with a poorly matched chromaticity $Q' \approx 20$}
\label{chroma20}
\end{center}
\end{figure}

Now, unfortunately, although the dispersion created in the dipole magnets requires nothing more than some more aperture in the vacuum chamber, the chromaticity of the quadrupoles has an influence on the tune of the particles and so can lead to dangerous resonance conditions. Particles with a particular momentum error will be pushed into resonances and be lost within a very short time.
A look at the tune spectrum visualizes the problem.  Whereas an ideal situation leads to a well-compensated chromaticity and the particles oscillate with basically the same frequency (Fig. \ref{chroma1}), a non-corrected chromaticity ($Q'$ = 20 units in the case of Fig.~\ref{chroma20})
broadens the tune spectrum and a number of particles are pushed towards dangerous resonance lines.

In large storage rings and synchrotrons in particular, this problem is crucial and represents one of the major factors that limit machine performance: because of the strong focusing of the quadrupoles and the large size, the  chromaticity can reach considerable values. A chromaticity correction scheme is therefore indispensable.
The trick is performed in three steps.
\begin{itemize}
\item We sort the particles in the horizontal plane according to their momentum.
This is done whenever we have a non-vanishing dispersion, for example close to the focusing quadrupoles in the arc, where both the dispersion and the $\beta$-function reach high values and the particle trajectories are determined by the well-known relation $x_\mathrm{d}(s) = D(s) \cdot \Delta p /p$.
\item At these places, we create magnetic fields that have a position-dependent focusing strength, in other words, fields that represent a position-dependent gradient. Sextupole magnets have exactly this property: if $g'$ describes the strength of the sextupole field, we get
\begin{equation}
B_x=g' \cdot xy
\end{equation}
for the horizontal field component and
\begin{equation}
B_y = g'\frac {1}{2} \cdot (x^2 - y^2 )
\end{equation}
for the vertical component.
The resulting gradient in both planes is obtained as
\begin{equation}
\frac{\rmd B_x}{\rmd y}=\frac{\rm d B_y}{\rmd x}= g'\cdot x .
\end{equation}
\item We now only have to adjust the strengths of two sextupole families (one to compensate the horizontal and another to compensate the vertical chromaticity) to get an overall correction in both planes.
\end{itemize}

In a little more detail, and referring again to normalized gradients, we can write
\begin{equation}
k_\mathrm{sext} = \frac{e}{p}g' \cdot x = m \cdot x ,
\end{equation}
which leads for a given particle amplitude
\begin{equation}
x_\mathrm{d} = D \cdot \frac {\Delta p}{p}
\end{equation}
to the normalized focusing strength (of the sextupole magnet)
\begin{equation}
k_\mathrm{sext} = m \cdot D \frac {\Delta p}{p} .
\end{equation}
The combined effect of the so-called natural chromaticity created by the quadrupole lenses (Eq.~(\ref{qprime})) and the compensation by the sextupoles leads to an overall chromaticity
\begin{equation}
Q' = - \frac{1}{4 \pi} \oint {(K(s)- m(s) \cdot D(s)) \beta(s)} \, \rmd s
\end{equation}
and needs to be compensated to zero in both transverse planes.

To summarize and make things as crystal clear as possible, the focusing properties of the magnet lattice lead to restoring forces in both transverse planes. The transverse motion of a particle is therefore a quasi-harmonic oscillation as the particle moves through the synchrotron, and the tune describes the frequency of these oscillations. As we cannot assume that all particles have exactly the same momentum, we have to take into account the effect of the momentum spread in the beam: the restoring forces are a function of the momentum of each individual particle and so the tune of each particle is different. We have to correct for this effect, and we do so by applying sextupole fields in regions where a non-vanishing dispersion distributes the off-momentum particles in the horizontal plane. \\
As easy as that!

\section{Longitudinal beam dynamics}

\subsection{Introduction}
Following the tradition in most textbooks, we treat particle dynamics in the longitudinal direction in a separate section of this paper. And, before going into the technical and mathematical details of the treatment, we would like to describe the motivation behind this decision.

Technically, magnetic fields are most appropriate for guiding the beam and creating the transverse focusing forces that are needed to keep the particles on a stable orbit. As we have shown in the discussion of Eq. (\ref{lorentz}), the electric fields that could be used as well are far weaker. However, the Lorentz force resulting from a magnetic field is always perpendicular to the particle's velocity vector. And, as an unfortunate consequence, we are obliged to make use of the weak electric fields as soon as we talk about particle acceleration. Technically speaking, we must replace our beloved magnets by electric fields, created in devices that we call RF resonators or cavities, where RF waves are built up to act on the beam, with the field vector pointing in the direction of the particle motion.

As a direct consequence, the resulting forces in the longitudinal plane are much weaker and so the longitudinal oscillation frequencies are much smaller. Thus, in addition to the technically quite different approaches to transverse and longitudinal dynamics, there is a physics argument that allows us to treat the two aspects independently. The resulting frequencies are very different, and there is hardly any crosstalk (or coupling) between the transverse and longitudinal oscillations of the beam particles. An extreme example might again be the LHC: whereas the frequency of the betatron oscillations is of the order of 1~MHz, the frequency of the longitudinal movements is about 23~Hz.
Nevertheless, for completeness, we should mention that in a higher order of approximation coupling between the two modes is indeed possible and needs to be avoided. But these topics are beyond the scope of this paper, and the curious reader is referred to Ref.~\cite{synchrobeta}.

The longitudinal movement of the particles in a storage ring --- and, closely related to this, the acceleration of the beam --- is strongly related to the problem of synchronization between the particles and the accelerating system. This synchronization may be established via the basic hardware and design of the machine (as in a Wideroe structure) or via the orbit (as in a cyclotron), or it may be a fundamental feature of the ring in a more sophisticated way, which leads to the name `synchrotron' for the specific type of machine concerned. We shall treat these different aspects in more detail. But, before we do so, we would like to start on a very fundamental basis and at the same time go back a little in history.

\subsection{Electrostatic machines}

\begin{figure}[h]
\begin{center}
\includegraphics[width=0.45\columnwidth]{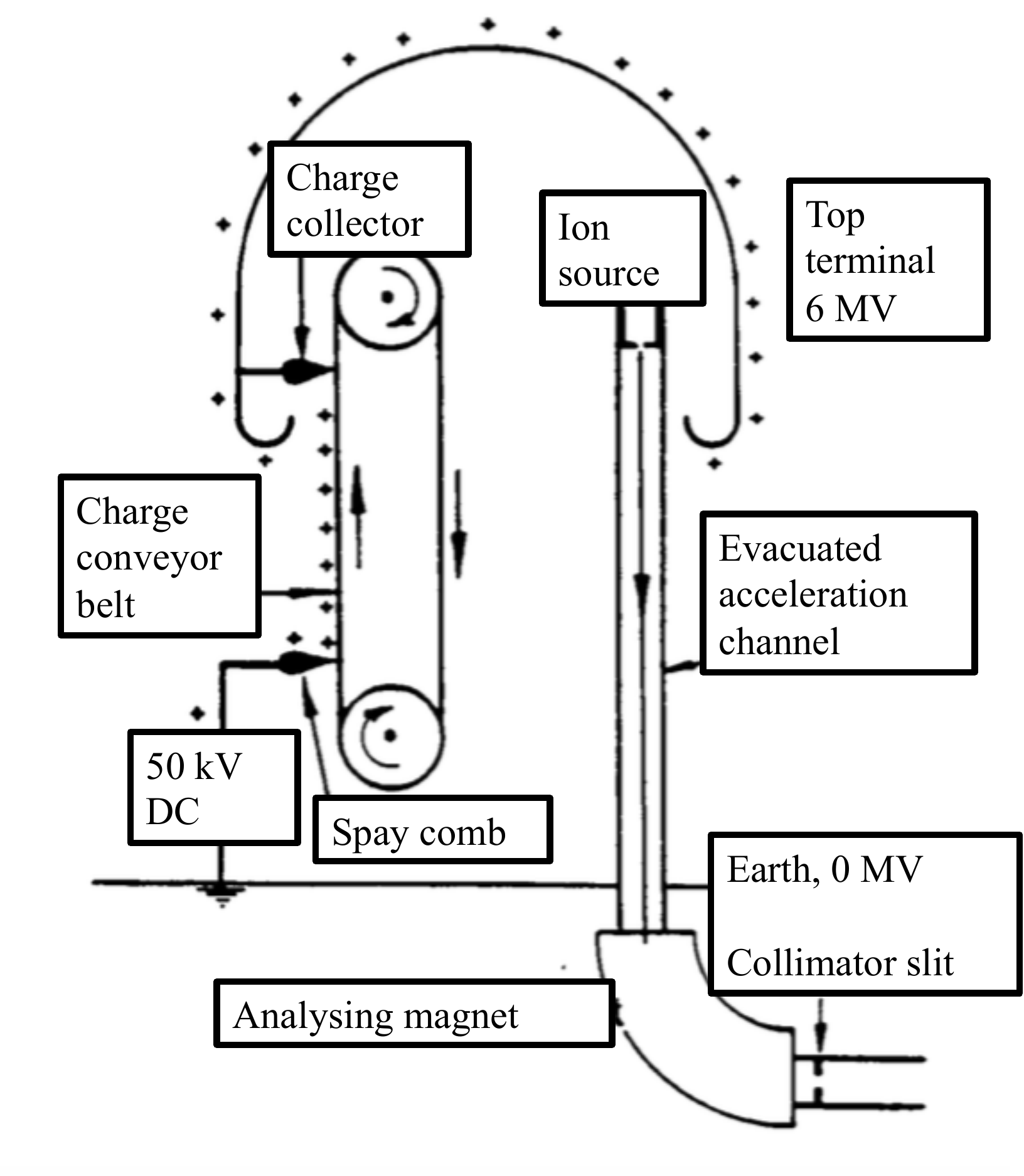}
\caption{Technical principle of a Van de Graaff accelerator \cite{Bryant}}
\label{vdg1}
\end{center}
\end{figure}

The most prominent example of an electrostatic machine, besides the Cockcroft--Walton generator \cite{Cockcroft1,Cockcroft2,Cockcroft3}, is the Van de Graaff accelerator \cite{vdGraaf}. A sketch of the principle is shown in Fig.~\ref{vdg1}.
Using a moderate DC high voltage, charges are sprayed onto a moving belt or a chain with insulated links and transported to a kind of Faraday screen, where the charges accumulate, leading to considerably high voltages. By their design concept, these machines deliver an excellent energy resolution, as basically each and every particle sees the same accelerating potential (which is no longer the case when we have to consider RF accelerating structures). A short overview of these machines can be found in Ref. \cite{Bryant}.

The kinetic energy of the particle beam is given by integration of the electric field $E$ in the direction $z$ of the particle motion, and is measured as usual in units of electronvolts (eV):
\begin{equation}
\mathrm{d}W = e \cdot E_z \, \mathrm{d}s \Rightarrow W = e \int E_z \, \rmd s = [\mathrm{eV}] .
\end{equation}

\begin{figure}[h]
\begin{center}
\includegraphics[width=0.55\columnwidth]{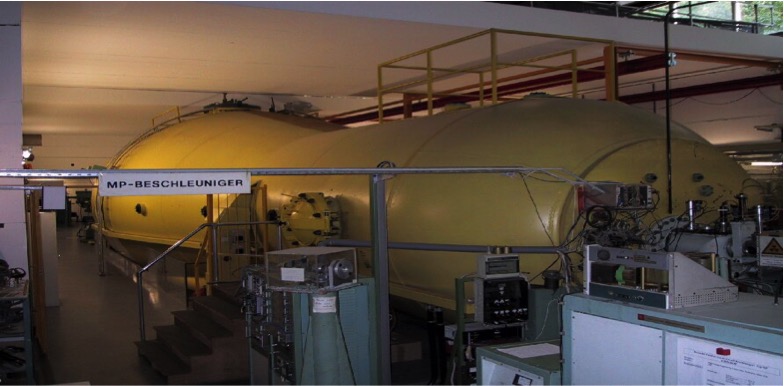}
\caption{Tandem Van de Graaff accelerator at the Max Planck Institute, Heidelberg}
\label{vdg2}
\end{center}
\end{figure}

An example of such a machine is shown in Fig.~\ref{vdg2}. This `tandem' Van de Graaff accelerator uses a stripper foil in the middle of the structure. Thus, in the first half, it accelerates negative ions that have been produced in a Cs-loaded source. After stripping of the electrons, the same voltage is applied in the second part of the structure to gain another step in energy. All in all, the particle energy is determined by the high voltage that can be created or, more precisely, by the potential difference the particles pass through, and so it is finally always limited by discharge effects in the electric field.

We emphasize here that  special synchronization (i.e., timing between the particles and accelerating field)  is not needed, as a DC voltage is used for acceleration. The disadvantage, however, is the fact that the DC voltage applied can only be used once and the energy gain of the particles is limited by the technically achievable voltage (or, more precisely, the feasible $E$-field), which defines the potential  difference.

\subsection{Radio frequency accelerators: The Wiederoe-type linac}
\begin{figure}[h]
\begin{center}
\includegraphics[width=0.70\columnwidth]{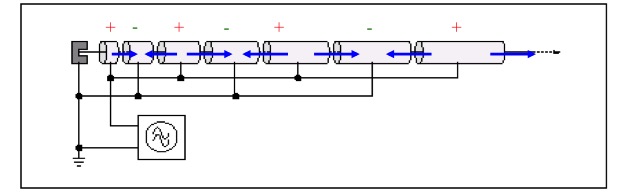}
\caption{Schematic view of a Wiederoe drift-tube linac}
\label{Wiederoe1}
\end{center}
\end{figure}

The  basic limitation imposed by the maximum achievable voltage in electrostatic accelerators can be overcome by applying AC voltages. However, a more complicated design is needed to prevent the particles from being decelerated during the negative half-wave of the RF system. In 1928, Wiederoe presented the layout of such a machine (and built it) for the first time. Figure \ref{Wiederoe1} shows the principle.

\begin{figure}[h]
\begin{center}
\includegraphics[width=0.55\columnwidth]{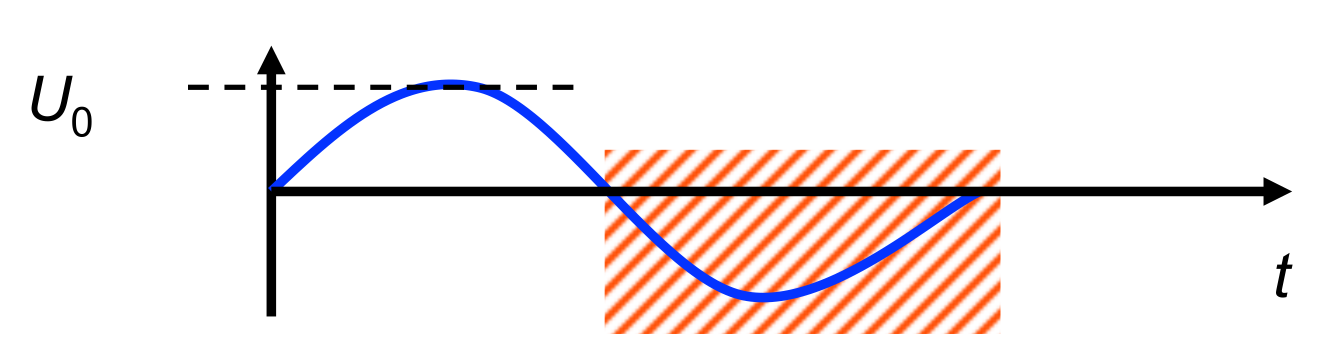}
\caption{The length of the drift tube has to be optimized to shield the particles from the negative RF half-wave}
\label{Wiederoe2}
\end{center}
\end{figure}

The arrows in the figure show the direction of the electric field at a given moment in time (i.e., the positive half-wave of the AC voltage), and the corresponding polarity applied to the electrodes is indicated by the $+$ and $-$ signs. In principle, arbitrarily high beam energies can be achieved by applying the same voltage to a high enough number of electrodes, provided that the particle beam is shielded from the electric field during the negative RF half-wave.
Accordingly, the electrodes are designed as drift tubes (thus the expression `drift tube linac') whose length is chosen according to the particle velocity and RF period. The design principle is shown schematically in Fig.~\ref{Wiederoe2}.

The area shaded red in the figure corresponds to the time during which the particle has to be shielded from the decelerating field direction, and is defined by the RF period: $t_\mathrm{shield} = \tau_\mathrm{rf}/2$. Accordingly, the length of the drift tube has to be
\begin{equation}
l_i = v_i \cdot \frac{\tau_\mathrm{rf}}{2} .
\end{equation}
If the kinetic energy (in the classical regime) is given by
\begin{equation}
E_i = \frac{1}{2}mv^2 ,
\end{equation}
we obtain an equation for the drift tube length that, for a given accelerating voltage $U_0$ and charge $q$, depends on the RF frequency $\nu_\mathrm{rf}$ and the number of the accelerating step $i$:
\begin{equation}
l_i=\frac{1}{\nu_\mathrm{rf}} \sqrt{\frac{iqU_0 \cdot \sin \psi_\mathrm{s}}{2m}} .
\end{equation}	
The parameter $\psi_\mathrm{s}$  describes the so-called synchronous phase and can be chosen to be $0^{\circ}$ in this case to obtain the highest acceleration performance.

\begin{figure}[h]
\begin{center}
\includegraphics[width=0.55\columnwidth]{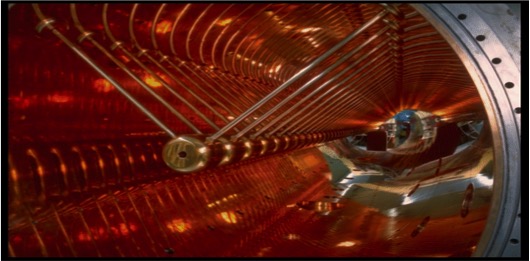}
\caption{The UNILAC drift tube accelerator at GSI}
\label{unilac}
\end{center}
\end{figure}

One of the best-known examples of such an accelerator is running at GSI in Darmstadt, and is used as a universal tool for the acceleration of (almost) any heavy-ion species. The internal structure is shown in Fig.~\ref{unilac}, including the drift tubes and the surrounding vessel.

Unlike the case for DC accelerators, the timing between the particles and the RF field is suddenly of major importance: synchronization has to be obtained between the particle's time of arrival at the resonator and the accelerating RF field, which---as shown above---is represented by the drift tube length, and so in a certain sense is built into the hardware of the system.

\section{Longitudinal particle dynamics in synchrotrons}
The design of a synchrotron follows this approach, except that the drift tube of the RF structure, where the particles are shielded from the decelerating field, is replaced by the machine itself.
If we define the longitudinal aspects of a synchrotron as a circular accelerator with:
\begin{itemize}
\item  a design orbit of constant radius, defined by the arrangement and strength of a number of dipole magnets;
\item  an RF system, located at a distinct place in the ring and powered at an RF frequency that is equal to the revolution frequency of the particles or an integer multiple (so-called harmonic) of it;
\end{itemize}
we are already quite close to reality. The rest is some mathematics.

\begin{figure}[h]
\begin{center}
\includegraphics[width=0.5\columnwidth]{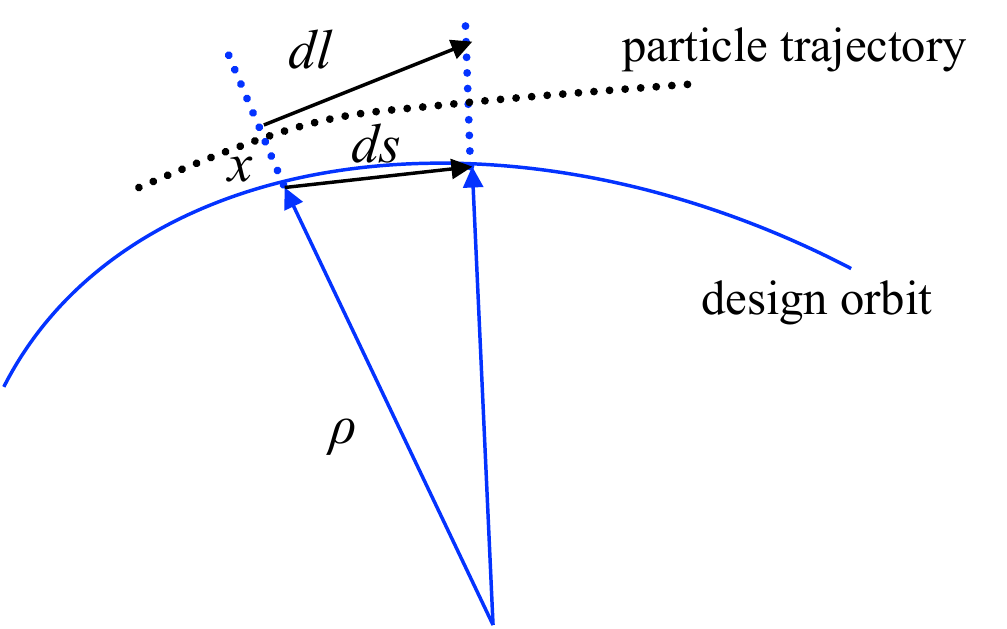}
\caption{Particle orbits in a synchrotron for an ideal and an off-momentum particle}
\label{dsdl}
\end{center}
\end{figure}

For a description of the particle dynamics, we refer to a synchronous particle of ideal energy, phase, and energy gain per turn. As we shall see, the synchronization between the RF system and the particle beam is of major importance in this type of machine and has to be explicitly included in the design. To understand the principle, we have to refer briefly to the transverse dynamics of a particle with a momentum error and the resulting dispersive effects (Fig.~\ref{dsdl}).

Whereas the ideal particle will run on the design orbit defined by the dipole magnets and will proceed a distance $\mathrm{d}s$, a non-ideal particle will run on a displaced orbit (displaced to the outer side of the ring in the example of Fig.~\ref{dsdl}) and will travel a corresponding distance $\mathrm{d}l$:
\begin{equation}
\frac{\rmd l}{\rmd s} = \frac{\rho+x} {\rho} .
\end{equation}	
Solving for $\mathrm{d}l$, we obtain
\begin{equation}
\rmd l = 1 + \frac{ x} {\rho(s)} \rmd s ,
\end{equation}	
and by integrating around the machine we get the orbit length of the non-ideal particle, which depends on the radial displacement $x$:
\begin{equation}
l_{\Delta E} = \int \rmd l = \int \left(1 +  \frac{ x_{\Delta E}} {\rho(s)} \right) \, \rmd s ,
\end{equation}	
where we assume that the radial displacement $x_{\Delta E}$ is caused by a momentum error and the dispersion function of the magnet lattice:
\begin{equation}
x_{\Delta E}(s) = D(s) \cdot \frac{\Delta p}{p} .
\end{equation}	
We obtain an expression for the difference in orbit length between the ideal and the dispersive particle, which is determined by the size of the relative momentum error and the dispersion function of the storage ring:
\begin{equation}
\delta l_{\Delta E} = \frac{\Delta p}{p} \int \frac{D(s)}{\rho(s)} \, \rmd s .
\label{dsdleq}
\end{equation}	
The ratio between the relative orbit difference and the relative momentum error is called the {\it momentum compaction factor} $\alpha_p$ and is determined by the integral of the dispersion function around the ring and the bending radius of the dipole magnets:
\begin{equation}
\frac{\delta l_{\Delta E}}{L}  = \alpha_p \frac{\Delta p}{p} ,
\label{momcomp_1}
\end{equation}	
where
\begin{equation}
\alpha_p = \frac {1}{L} \int \frac{D(s)}{\rho(s)} \, \rmd s .
\label{momcomp}
\end{equation}	

Although the expression `momentum compaction factor' might be an unfortunate choice and we would like instead to call it the lengthening factor, its physical meaning is as important as it is clear: it describes the lengthening of the orbit for particles that have a given momentum deviation with respect to the ideal particle. And it is also clear that in a circular accelerator this orbit-lengthening effect cannot be avoided, because of the dispersion function.

For some initial estimates, we assume equal bending radii in all dipoles, so $1/\rho = \mathrm{const}$ and we replace the integral of the dispersion around the ring by a sum over the average dispersion in the dipole magnets (outside the dipoles the term $1/\rho =0$, so this assumption is justified for a rough estimate):
 \begin{equation}
\int_\mathrm{dipoles} D(s) \, \rmd s \approx l_{\Sigma (\mathrm{dipoles})} \langle D \rangle _\mathrm{dipole} .
\end{equation}	
We get a nice, simple expression for the momentum compaction factor that depends only on the ratio of the average dispersion to the geometric radius $R$ of the machine:
\begin{equation}
\alpha_p = \frac {1}{L}  l_{\Sigma (\mathrm{dipoles})} \langle D \rangle _\mathrm{dipole} \frac {1}{\rho} = \frac{1}{L}2\pi \rho \langle D \rangle \frac{1}{\rho} ,
\end{equation}	
\begin{equation}
\alpha_p = \frac{2\pi}{L} \langle D \rangle \approx \frac{\langle D \rangle}{R} .
\end{equation}	
For a quick estimate, $\alpha_{p}$ is given by the ratio of the average dispersion to the geometric radius of the ring.
Assuming, finally, that the particles are moving at the speed of light, i.e., $v \approx c = \mathrm{const} $, the relative error in time 
is given by the relative change in the orbit length and thus by the momentum compaction factor and the relative momentum error:
\begin{equation}
\frac{\delta t}{t} = \frac{\delta l_{\epsilon}}{L}=\alpha_p \frac{\Delta p}{p} .
\label{eq54}
\end{equation}	

So the secret of the longitudinal motion is already disclosed, 
even if you might not have realized yet: the dispersive effect in a synchrotron or, in other words, the orbit lengthening for off-momentum particles, described by $\alpha_p$, is the fundamental feature of the principle of operation of a synchrotron: it relates the time of arrival in the RF structure to the momentum error of the particle.

\subsection{Dispersive effects in synchrotrons}

Well \ldots we have seen this topic come up already, but we would like to look at the issue from the point of view of timing, i.e., from the point of view of the synchronization between the particles and the RF system. And we have to enlarge our point of view and include the case of non-relativistic particles (a rigorous treatment can be found, for example, in Refs. \cite{tecker} and \cite{joel}).  So our problem of synchronization needs a more careful treatment, which must include the fact that the particles are travelling at a speed that might be considerably lower than the speed of light. The parameter of interest, however, is still the ratio between the relative momentum error and the relative frequency deviation of a particle:
\begin{equation}
\frac{\rmd  f_\mathrm{r}}{f_\mathrm{r}} = \eta  \frac{\rmd p}{p} .
\end{equation}	
Here we have introduced the parameter $\eta$ to address this issue, and we have added explicitly a subscript `r' to denote the revolution frequency of the particle.
We shall derive an expression for this $\eta$-parameter in the next few lines. But we would like to point out now that $\eta$ combines the effect of the changing velocity of the particle {\it and} the relativistic increase  in mass with changing energy. And thus it is this strange $\eta$ that is the key parameter for anything about timing in a synchrotron.

But step by step $\dots$ Given the revolution frequency as a function of machine circumference and speed, the revolution frequency around the ring is
\begin{equation}
 f_\mathrm{r}  =  \frac{\beta c }{2 \pi R} ,
\end{equation}	
where $\beta$ is the relativistic parameter $v/c$ and $R$ stands for the geometric radius of the machine (defined by the length of the design orbit, which is $2 \pi R$). Via the logarithmic derivative, we obtain the obvious relation
\begin{equation}
 \frac{ \rmd f_\mathrm{r}}{f_\mathrm{r}}  =  \frac{\rmd \beta }{\beta} - \frac{\rmd R}{R} .
 \label{frbetaR}
\end{equation}	
Now, from Eq.~(\ref{momcomp_1}) we know that the relative change in radius, i.e., the second term in the expression, is given by the momentum compaction factor $\alpha_p$:
\begin{equation}
 \frac{\rmd R}{R}=\alpha_p \frac{\rmd p}{p} ;
 \label{dr_r}
\end{equation}	
and, as the momentum is related to the particle energy,
\begin{equation}
p=mv=\beta \gamma \frac{E_0}{c},
\end{equation}	
we can write the following for the relative momentum change:
\begin{equation}
\frac{\rmd p}{p} = \frac{\rmd \beta}{\beta} + \frac{\rmd (1-\beta^2)^{-1/2}}{(1-\beta^2)^{-1/2}} = \gamma^2 \frac{\rmd \beta}{\beta} .
\label{dp_p}
\end{equation}	

Introducing the two equations (\ref{dr_r}) and (\ref{dp_p}) into Eq.~(\ref{frbetaR}), we finally obtain the required relation between the frequency offset and the momentum error,
\begin{equation}
 \frac{\rmd f_\mathrm{r}}{f_\mathrm{r}}  =  \left(\frac{1}{\gamma^2} - \alpha \right) \frac{\rmd p}{p} .
 \label{xxx}
 \end{equation}	
Accordingly, the $\eta$-parameter defined above is given as
\begin{equation}
\eta = \frac{1}{\gamma^2}-\alpha_p .
\end{equation}	
This combines the effect of a momentum deviation on the orbit size (described by $\alpha_p$ above) and the effect of the speed of the particle, which increases with increasing momentum, until we reach the ultrarelativistic regime and $v \approx c = \mathrm{const}$.

This relation is indeed extremely interesting, as it tells us {\it what to do when we start to accelerate particles} in a synchrotron.
The easiest situation occurs when we are dealing with ultrarelativistic particles. In this case the last equation reduces to the simplified situation
described in Eq.~(\ref{eq54}): if $\gamma$ is high,
the first term, $1/\gamma^2$, tends to zero and the change in the revolution frequency is defined by the momentum compaction factor $\alpha_p$.
For small energies, however, things get more complicated.

As an important remark, we state that the change in revolution frequency depends on the particle energy $\gamma$ and may possibly change sign during acceleration. Particles become faster at the beginning of the process and arrive earlier at the location of the cavity (classical regime), whereas particles that travel at $v \approx c$ will not get any faster but instead become more massive and, being pushed to a dispersive orbit, will arrive later at the cavity (relativistic regime). The boundary between the two regimes is defined by the case where no dependence of the frequency on $\rmd p/p$ is obtained, namely, $\eta = 0$, and the corresponding energy is called the {\it transition energy}:
\begin{equation}
\gamma_\mathrm{tr} = \frac{1}{\sqrt \alpha_p} .
\end{equation}	

\begin{figure}[h]
\begin{center}
\includegraphics[width=0.7\columnwidth]{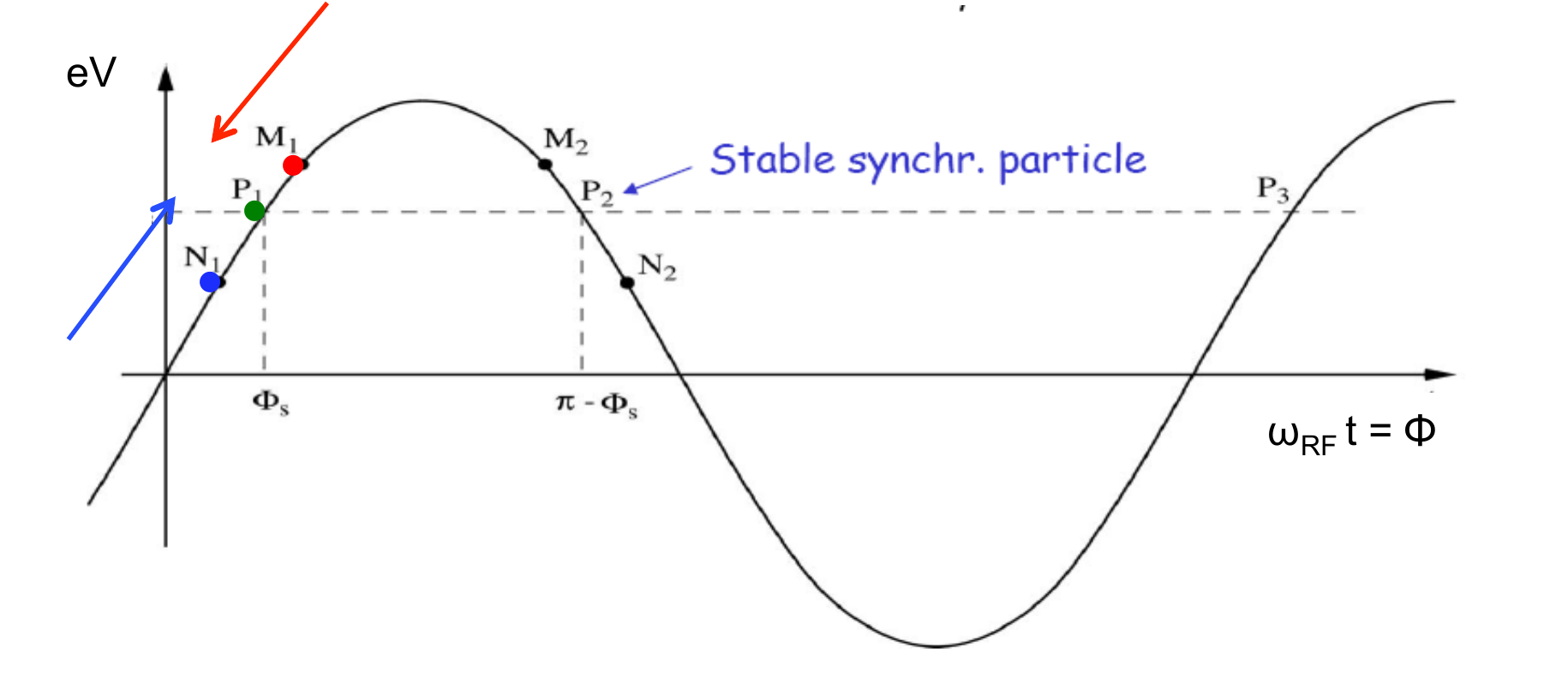}
\caption{Qualitative picture of the phase-focusing principle below the transition}
\label{synch_1}
\end{center}
\end{figure}

In general, we design machines in such a way as to avoid the crossing of this transition energy. As it involves changes in the RF phase unless the particles lose the longitudinal focusing created by the sinusoidal RF function, the bunch profile will be diluted and become lost. Qualitatively, the longitudinal focusing effect and the problem of the gamma transition are explained in Fig.~\ref{synch_1}.

\subsection{The classical regime}
Assume that an ideal particle is passing through the cavity at a certain ideal position in time (or phase), as indicated by the green spot in Fig.~\ref{synch_1}. 
It will see a certain accelerating voltage and, correspondingly, receive an energy increase. We call this phase the synchronous phase. A particle that has a smaller energy than the ideal value will travel at a lower speed and will arrive later after the next turn, and thus at a larger phase, and it will see a stronger accelerating voltage. It will therefore compensate the lack in energy and, step by step, come closer to the ideal particle. Just the opposite happens to a particle that has a positive energy offset. As it is faster than the synchronous particle, it will arrive at the cavity earlier and see a smaller voltage, and will again approach the ideal particle step by step. In both cases a net focusing effect is obtained, which is due to the relation between momentum and speed and the right choice of the synchronous phase. This focusing effect leads to stable longitudinal oscillations of the particles, keeping them close together or, more precisely, close to the synchronous particle, and so it forms a so-called bunch of particles in the longitudinal direction.

Here we have to pause for a moment and contemplate the situation a little: it is evident that perfect synchronization can be obtained in the case where the revolution frequency $f_\mathrm{r}$ is equal to the RF frequency $f_\mathrm{rf}$. But it is also evident that we can again obtain a synchronous  condition if $f_\mathrm{rf} $ is an integer multiple of $f_\mathrm{r}$:
\begin{equation}
f_\mathrm{rf} = h  f_\mathrm{r} .
\end{equation}	
We call the integer $h$ the `harmonic number', and it defines the number of synchronous 'locations' on the closed orbit. This is clear enough: at each of these $h$ locations, the longitudinal-focusing principle is equally valid. Therefore we obtain $h$ so-called `buckets' in the machine that can be occupied by particle bunches.

\begin{figure}[h]
\begin{center}
\includegraphics[width=0.7\columnwidth]{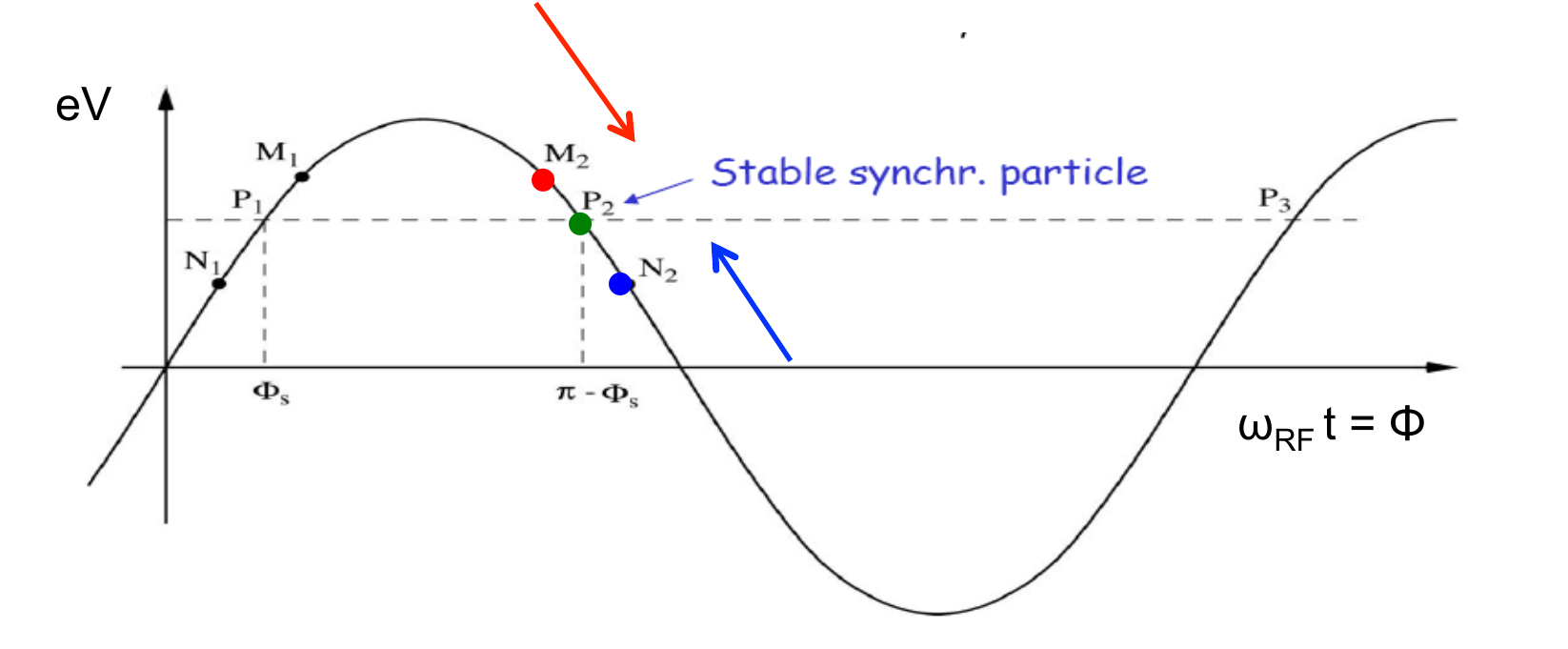}
\caption{Qualitative picture of the phase-focusing principle above the transition}
\label{synch_2}
\end{center}
\end{figure}

For highly relativistic particles, the same effect exists but the origin of the focusing effect is now the relativistic increase in mass with energy. As visualized in Fig.~\ref{synch_2}, the high-energy particle (marked in blue) will, because of its higher mass, move on a longer orbit and, as its speed is constant ($v \approx c $), it will arrive later at the cavity location. As a consequence, the synchronous phase has to be chosen depending on whether we are running the machine below or above the transition. Synchrotrons that have to pass through the transition will have to apply a phase jump to keep the particles bunched.

In this context, it is worth taking a look at the acceleration mechanism itself. The particle momentum is defined via the beam rigidity by the dipole field $B$:
\begin{equation}
p=e B \rho .
\end{equation}	
As a consequence, a change in the particle momentum is reflected by an appropriate change in the $B$-field:
\begin{equation}
\frac{\rmd p}{\rmd t}=e \rho \dot{B} .
\end{equation}	
The momentum increase per turn is therefore given by
\begin{equation}
(\Delta p)_\mathrm{turn} \, {\rmd t} = e \rho \dot{B} T_\mathrm{r} = \frac{2 \pi e \rho R  \dot{B}}{v}
\end{equation}	
and, referring to the energy change rather than the change in momentum, we obtain using
\begin{equation}
E^2 = E_0^2 +p^2 c^2 \rightarrow \Delta E= v \, \Delta p
\end{equation}	
the change in energy per turn, which is clearly related to the accelerating voltage and the synchronous phase $\phi_\mathrm{s}$  of the particles:
\begin{equation}
\Delta E_\mathrm{turn} = \Delta W_\mathrm{turn} = 2 \pi e \rho R \dot{B} = e \hat{V} \sin \phi_\mathrm{s} .
\end{equation}	

The following remarks might be worth making.
\begin{itemize}
\item The dipole field changes the orbit, and this leads to a change in the time (or phase) of arrival at the RF cavities and so to an accelerating effect on the whole beam.
\item The energy gain depends on the rate of change of the dipole field.
\item The number of stable synchronous particles is equal to the harmonic number $h$, which is ultimately the number of RF wavelengths that fit into the machine circumference. Thus we get $h$ synchronous particles that are equally spaced around the circumference.
\item All synchronous particles satisfy the relation $p = eB\rho$. They have the nominal energy and follow the nominal trajectory.
\item As long as the particles are not fully relativistic, their revolution frequency changes, and so the RF frequency, which is a multiple of $f_\mathrm{r}$, must also change to stay in synchronization during the complete acceleration process.
\end{itemize}

\subsection{Frequency change during acceleration}
One last comment might be useful before we can go to our workshop, take a jigsaw and hammer, and start building a machine.
As soon as we start to accelerate a particle in our ring, two things will happen according to Eq.~(\ref{frbetaR}): the velocity will increase (which  justifies the term `accelerator') and, at the same time, the relativistic mass will increase via $m=\gamma \cdot m_0$. Because of the first effect, we expect a change in revolution frequency, and because of the second this velocity change will reduce and the mass effect will take over.

Now, the relation between the revolution frequency and the RF frequency is defined by the harmonic number and depends on the size of the ring and the magnetic dipole field:
\begin{equation}
f_\mathrm{r} = \frac{f_\mathrm{rf}}{h} = \mathrm{function} (B, R_\mathrm{s}) .
\end{equation}	
Hence, using the beam rigidity relation and the average dipole field to define the radius of the ideal particle, we obtain the following for an average magnetic field $\langle B(t) \rangle $:
\begin{equation}
\frac{f_\mathrm{rf}(t)}{h} = \frac{v(t)}{2 \pi R_\mathrm{s}} = \frac{1}{2 \pi} \frac{e}{m} \langle B(t) \rangle
\end{equation}	
and
\begin{equation}
\frac{f_\mathrm{rf}(t)}{h} = \frac{1}{2 \pi} \frac{ec^2}{E_\mathrm{s}(t)} \frac{r}{R_\mathrm{s}} B(t) .
\end{equation}	

I hope it is becoming clearer that the `independent parameter' that drives the particle acceleration in a synchrotron is the magnetic dipole field
(even if we have to admit that a certain number of RF cavities is also useful for doing the job).
Using the relativistic overall energy
\begin{equation}
E^2 = p^2c^2+(m_0c^2)^2 ,
\end{equation}	
we finally obtain an expression for the RF frequency as a function of the changing external dipole field:
\begin{equation}
\frac{f_\mathrm{rf}(t)}{h} = \frac{c}{2 \pi R_\mathrm{s}} \left\{ \frac{B(t)^2}{(m_0 c^2 /ecr)^2+B(t)^2} \right\}^{1/2} .
\label{frf_B}
\end{equation}	

So all we have to do for successful operation is to put the form of our ramping $B$-field into  Eq.~(\ref{frf_B}) and adjust the
frequency control of our RF system accordingly. And the machine will run automatically and be `synchronized'---which is where the name ultimately comes from.

At high energies, or, more accurately, when
\begin{equation}
B > \frac{m_0c^2}{ecr} ,
\label{74}
\end{equation}	
the velocity increase becomes increasingly small (we get closer and closer to the speed of light), the second term becomes negligible, and the situation simplifies a quite a bit:
\begin{equation}
\frac{f_\mathrm{rf}(t)}{h} = \frac{c}{2 \pi R_\mathrm{s}} = \mathrm{const} .
\label{75}
\end{equation}	
In the case of electron synchrotrons, because of the small mass of electrons and, as a consequence, the high values of $\gamma$, it is evident
that the condition for Eq.~(\ref{75}) is nearly always fulfilled and that that relation can be applied right from the beginning: in these machines, the revolution frequency does not change by any considerable amount during acceleration and can be considered as constant.
For proton and heavy-ion beams, however, even up to LHC energies ($7$~TeV in the case of protons), the effect has to be taken into account up to the flat-top energy
and so proton and heavy-ion synchrotrons need more sophisticated RF control.

\section{Synchrotron motion}

\begin{quotation}
Once more unto the breach, dear friends \ldots \cite{shakespear}
\end{quotation}
In the following, we shall again contemplate the longitudinal motion a little. However, we shall try to put things on a mathematically more solid basis. As shown qualitatively in Fig.~\ref{synch_2},  we expect a longitudinal oscillation in phase and energy under the influence of the focusing mechanism explained above.
The relation between the relative frequency deviation and relative momentum error has been derived in (Eq.~(\ref{xxx})):
\begin{equation}
 \frac{\rmd f_\mathrm{r}}{f_\mathrm{r}}  =  \left(\frac{1}{\gamma^2} - \alpha_{p} \right) \frac{\rmd p}{p} ,
 \end{equation}	
which translates into a difference in revolution time
\begin{equation}
 \frac{ \rmd T}{T_0} = \left(\alpha_{p} - \frac{1}{\gamma^2} \right) \frac{\rmd p}{p}
 \end{equation}	
and leads to a difference in phase on arrival at the cavity
\begin{eqnarray*}
\Delta \psi  & = & 2 \pi \frac{\Delta T}{T_\mathrm{rf}} = \omega_\mathrm{rf} \cdot \Delta T
\\
 & = & \frac{h \cdot 2 \pi}{\beta^2} \left(\alpha_p - \frac{1}{\gamma^2} \right) \frac{\rmd E}{E}
\\
 & = & h \cdot \omega_0 \cdot \Delta T = h 2 \pi \frac{\Delta T}{T_0}
\\
 & = & h \cdot 2 \pi \left(\alpha_p - \frac{1}{\gamma^2} \right) \frac{\rmd p}{p}
\\
 & = & \frac{h \cdot 2 \pi}{\beta^2} \left(\alpha_p - \frac{1}{\gamma^2} \right) \frac{\rmd E}{E} .
\label{test_mult_eq}
\end{eqnarray*}
As before, the revolution frequency $f_\mathrm{r}$ and the RF frequency $\omega_\mathrm{rf}$ are related to each other via the harmonic number $h$. Hence the difference in energy and the offset in phase are connected to each other through the momentum compaction factor or, more accurately, the parameter $\eta$.

Differentiating the last expression with respect to time gives the rate of change of the phase offset per turn:
\begin{equation}
\Delta \dot{\psi} = \frac{\Delta \psi}{T_0} = \frac{h 2 \pi}{\beta^2 T_0} \left(\alpha_p - \frac{1}{\gamma^2} \right) \frac{\rmd E}{E} .
\label{psidot}
\end{equation}	
This expression tells us about the rate of change of the phase of a particle as a function of its changing energy.

On the other hand, the difference in energy gain of an arbitrary particle that has a phase distance
of $\Delta \psi$ from the ideal particle is given by the voltage and phase of the RF system (a trivial statement but worth mentioning):
\begin{equation}
\Delta E = e \cdot U_0 ( \sin(\psi_\mathrm{s} + \Delta \psi) -\sin \psi_\mathrm{s} ) .
\end{equation}	
As before, we describe the phase of the ideal (`synchronous') particle by $\psi_\mathrm{s}$ and the phase difference by $\Delta \psi$. For small amplitudes $\Delta \psi$ of the phase oscillations, we can simplify the treatment by assuming
\begin{equation}
 \sin(\psi_\mathrm{s} + \Delta \psi) - \sin \psi_\mathrm{s} = \sin \psi_\mathrm{s} \cos \Delta \psi -  \cos \psi_\mathrm{s} \sin \Delta \psi - \sin \psi_\mathrm{s} .
\end{equation}
For small amplitudes $\Delta \psi$, we can make the approximation
\begin{equation}
 \sin \Delta \psi \approx \Delta \psi, \hspace{0.5cm}   \cos \Delta \psi \approx 1 ,
\end{equation}
and obtain the following for the rate of energy change per turn:
\begin{equation}
\Delta \dot{E} = e \cdot \frac{U_0}{T_0}  \Delta \psi \cos \psi_\mathrm{s} .
\end{equation}	
A second differentiation with respect to time delivers
\begin{equation}
\Delta \ddot{E} = e \cdot \frac{U_0}{T_0}  \Delta \dot{\psi}  \cos \psi_\mathrm{s} .
\label{Edotdot}
\end{equation}	
Combining Eqs.~(\ref{psidot}) and (\ref{Edotdot}), we finally get a differential equation for the longitudinal motion under the influence of the phase-focusing mechanism:
\begin{equation}
\Delta \ddot{E} = e \cdot \frac{U_0}{T_0}  \frac{2 \pi h}{\beta^2 T_0} \left(\alpha_p - \frac{1}{\gamma^2} \right) \frac{\rmd E}{E} \cos \psi_\mathrm{s} .
\end{equation}	
For a given energy, the parameters in front of the right-hand side are constant and describe the
longitudinal, or `synchrotron', oscillation frequency. Therefore, using
\begin{equation}
\Omega = \omega_0 \cdot \sqrt{\frac{-eU_0 h \cos \psi_\mathrm{s}}{2 \pi \beta^2 E} \left(\alpha_p - \frac{1}{\gamma^2} \right)} ,
\label{Omega}
\end{equation}	
we get the equation of motion in the approximation of small amplitudes:
\begin{equation}
\Delta \ddot{E} + \Omega^2 \, \Delta E = 0 .
\end{equation}	
This describes a harmonic oscillation in ($E$--$\psi$) phase space of the difference in energy of a particle from the ideal (i.e., synchronous) particle under the influence of the phase-focusing effect of our sinusoidal RF function.

\begin{figure}[h]
\begin{center}
\includegraphics[width=0.38\columnwidth]{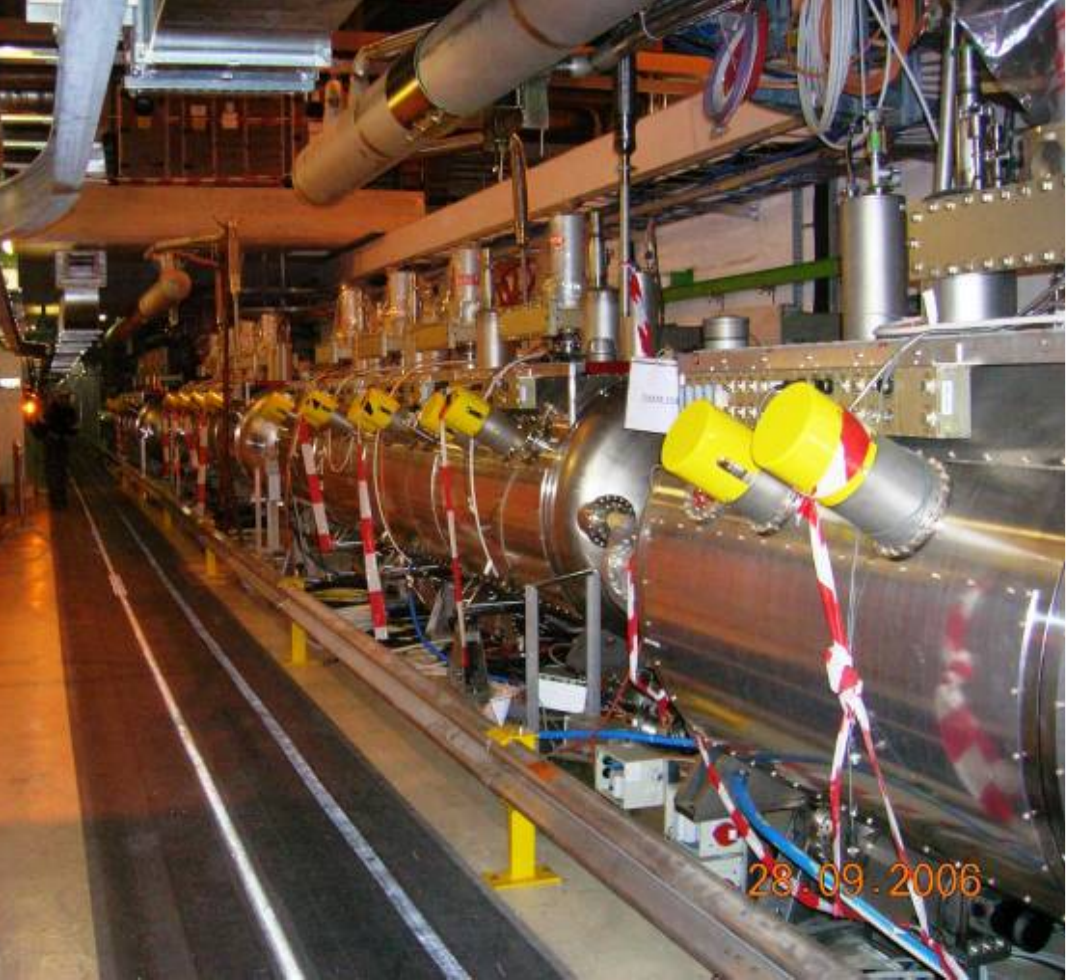}
\caption{LHC RF system}
\label{lhcrf}
\end{center}
\end{figure}

As already discussed qualitatively, the expression in Eq.~(\ref{Omega}) leads to real solutions if the argument of the square root is a positive number. Two possible situations therefore have to be considered: below the gamma transition the $\eta$-parameter is positive, and above it it is negative. The synchronous phase (which is the argument of the cosine function) therefore has to be chosen to get an overall positive value under the square root:
\begin{eqnarray*}
\gamma < \gamma_\mathrm{tr} \hspace{1cm} \eta > 0, \hspace{0.5cm} 0<\psi_\mathrm{s} < \pi/2 , \\
\gamma > \gamma_\mathrm{tr} \hspace{1cm} \eta < 0, \hspace{0.5cm}  \pi/2 < \psi_\mathrm{s} < \pi .
\end{eqnarray*}	
This, finally, is the mathematical description of the classical and relativistic regimes that defines stable conditions of synchrotron motion, as explained qualitatively in Figs.~\ref{synch_1} and \ref{synch_2}.
Figure \ref{lhcrf} shows as an example the superconducting RF system of the LHC, and the basic parameters of this system, including the synchrotron frequency, are listed in Table \ref{lhc_rf_param}.

\begin{table}[h]
\caption{Parameters of the LHC RF system}
\centering
  \begin{tabular}{ l c }
    \hline\hline
Bunch length ($4\sigma$)     	             	&  1.06~ns                      \\
Energy spread ($2\sigma$)    	             	&   $0.22 \times 10^{-3}$              	\\
Number of stored bunches                      	&   2808                     \\
RF frequency 	  			                    &  400~MHz                       \\
Harmonic number $h$			     				&   335\,640                \\
RF voltage per beam                				&    16~MV                       \\
Energy gain per turn      		    			&     485~keV                    \\
Synchrotron frequency  		         			&  23~Hz  			\\
 \hline\hline
  \end{tabular}
\label{lhc_rf_param}
\end{table}

\begin{figure}[h]
\begin{center}
\includegraphics[width=0.63\columnwidth]{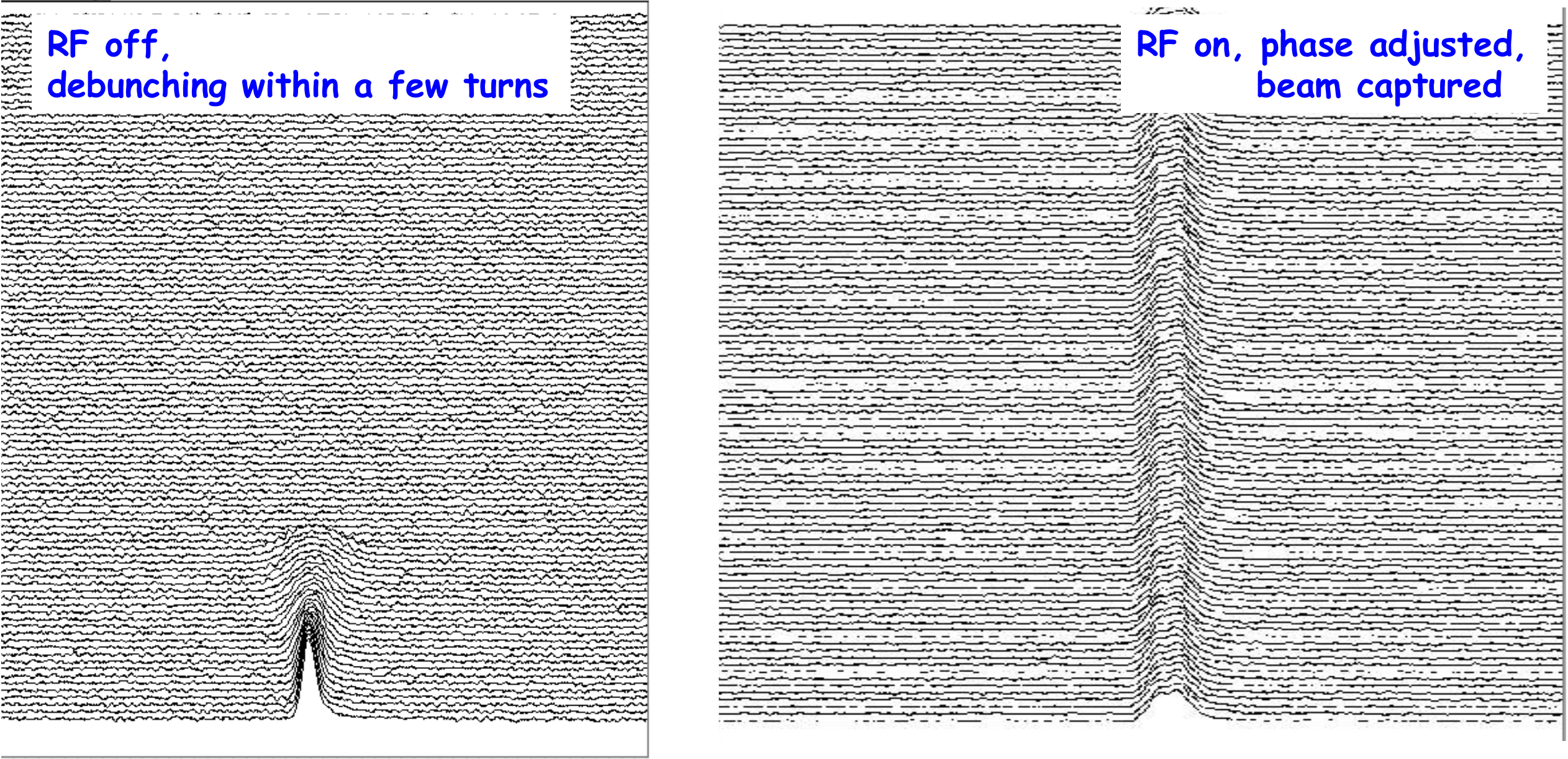}
\caption{Left: injection into the LHC while the RF is switched off: the bunches decay in a small number of turns.
Right: the RF is switched on and phased: the bunches remain long, focused, and stable.}
\label{mountain_range_1}
\end{center}
\end{figure}

Figure \ref{mountain_range_1}, finally, shows the longitudinal focusing effect that was observed during the commissioning phase of the LHC. On the left-hand side, beam had been injected into the storage ring while the RF system still was switched off. The bunch, nicely formed by the RF voltage of the pre-accelerator, is visible for only a few turns and the bunch profile decays rapidly, as no longitudinal focusing is active. The right-hand side shows the situation with the RF system activated and the phase adjusted. The injected particles stay nicely bunched and the acceleration process can start.


\newpage

\bibliography{bibliography/converted_to_latex.bib%
}

\end{document}